\documentclass[twocolumn]{aastex701}

\newcommand{\cm}{cm$^{-1}$}
\newcommand{\mn}{\ion{Mn}{1}}
\newcommand{\mnii}{\ion{Mn}{2}}

\usepackage{gensymb}
\usepackage{rotating}
\usepackage{multirow}
\usepackage{array}
\usepackage{tablefootnote}
\usepackage{nicefrac}
\usepackage{longtable}

\begin{document}

\title{Wavelengths and Energy Levels of Neutral Manganese (\ion{Mn}{1}) Determined Using High-Resolution Fourier Transform and Grating Spectroscopy}

\correspondingauthor{Christian Clear}
\email{christian.clear@imperial.ac.uk}

\author[0000-0002-3339-7097]{Christian P. Clear}
\affiliation{Physics Department, Imperial College London, London, SW7 2AZ, UK}
\email{christian.clear@imperial.ac.uk}

\author[0000-0002-0786-7307]{Gillian Nave}
\affiliation{Physics Department, Imperial College London, London, SW7 2AZ, UK}
\affiliation{National Institute of Standards and Technology, Gaithersburg, MD, 20889-8422, USA}
\email{fake@gmail.com}

\author{Richard Blackwell-Whitehead}
\affiliation{Physics Department, Imperial College London, London, SW7 2AZ, UK}
\email{fake@gmail.com}

\author[0000-0002-3527-201X]{Maria Teresa Belmonte}
\affiliation{Physics Department, Imperial College London, London, SW7 2AZ, UK}
\affiliation{Universidad de Valladolid, Departamento de F\'isica Te\'orica, At\'omica y \'Optica, P. de Bel\'en, 7, Valladolid 47011, Spain}
\email{fake@gmail.com}

\author{Stephen Ingram}
\affiliation{Physics Department, Imperial College London, London, SW7 2AZ, UK}
\email{fake@gmail.com}

\author[0000-0003-2879-4140]{Juliet C. Pickering}
\affiliation{Physics Department, Imperial College London, London, SW7 2AZ, UK}
\email[]{j.pickering@imperial.ac.uk}



\begin{abstract}

An extensive analysis of the spectrum of neutral manganese has been performed using spectra of manganese-neon and manganese-argon hollow cathode discharges measured using high resolution Fourier transform (FT) and grating spectroscopy over the range 151 - 5112~nm (1956 - 65876 cm$^{-1}$). Wavelengths for 10426 spectral lines were extracted from the FT spectra, with uncertainties at least an order-of-magnitude lower than previous measurements. Wavelengths for 13397 lines from new grating spectra were  determined for spectral regions beyond the FT spectra range or to provide wavelengths for weak transitions not observed in FT spectra. To aid in level identification, selected, previously published grating lines were included in the energy level optimisation, but no levels in this work relied solely on previously published wavelengths. In total, 24237 lines were included in the final spectral linelist, and these were used to identify 2186 {\mn} transitions. These classified spectral lines were then used to optimise the values of 384 previously published energy levels of {\mn}, with typical uncertainties of a few $10^{-3}$ {\cm}, again typically an order-of-magnitude improvement in accuracy. Our study then expanded the known energy level structure of {\mn} through the establishment of 18 new energy levels, reported here for the first time. In total, 2187 lines and 402 energy levels of {\mn} have been determined as a result of our work, marking a substantial advance in the precision of {\mn} atomic data which will enable far more accurate analyses of {\mn} lines in astrophysical spectra.

\end{abstract}



\section{Introduction}
\label{sect:intro}

The rapid advancement of high-resolution spectrographs on both ground-based and space-borne telescopes, such as HST-STIS, VLT-UVES, and JWST, has dramatically expanded our ability to observe astrophysical spectra in unprecedented detail. However, the analysis of these spectra continues to be hindered by a lack of sufficiently accurate and comprehensive laboratory data, particularly among the iron-group elements. These elements, with their complex electron configurations and high relative abundances, give rise to dense forests of spectral lines that are prominent across the ultraviolet (UV) to infrared (IR) regions, playing a critical role in shaping the absorption and emission features of many astrophysical environments. In response, collaborative efforts led by Imperial College London (ICL) and the National Institute for Standards and Technology (NIST) have focussed on improving wavelengths and energy levels for neutral species through high-resolution Fourier transform (FT) and grating spectroscopy. These measurements are essential for interpreting stellar spectra, especially in cool stars and early-type stellar atmospheres, where neutral atomic species dominate spectral opacities.

Modern stellar astrophysical models are now being analysed in three dimensions and are taking into account departures from local thermodynamic equilibrium. Non-local thermodynamic equilibrium (NLTE) conditions arise when atomic level populations are determined by radiation fields and non-local processes, rather than being governed solely by the local plasma temperature and density. Neutral manganese ({\mn}) is particularly sensitive to NLTE effects, more so than {\mnii} \citep{Bergemann2019}, where effects such as overionisation, driven by ultraviolet radiation, can significantly weaken spectral lines, leading to systematic underestimation of Mn abundances if LTE is assumed. Modern stellar models require large amounts of atomic data to enable reliable NLTE modelling and improve abundance determinations across a wide range of stellar environments, underscoring the urgent need for high-precision atomic data for {\mn}, particularly accurate energy levels and transition wavelengths.

\subsection{Previous Measurements of {\mn}}
\label{sect:prev_measurements}
Previous measurements of the spectrum of {\mn} fall into two main groups. The older group consists of low resolution spectral line measurements obtained from broad wavelength range spectra, conducted using grating and prism spectrometers. The majority of energy levels derived from these older line measurements have uncertainties of $\sim$0.1 {\cm}. These data now fall far below the accuracy requirements of modern, high resolution astrophysical spectroscopy. The more recent group of measurements have been made using laser spectroscopy and interferometric techniques, typically with higher resolutions (0.05 {\cm} to 0.0001 {\cm}). However, these newer measurements have been for specific, limited sets of lines, accounting for a small fraction of the total observable transitions in the spectrum of {\mn} and are primarily concerned with studies of hyperfine structure (HFS) or the determination of atomic energy level lifetimes.

The first large-scale study of the spectrum and energy level system of {\mn} was performed by \citet{Cataln1923} using prism and grating spectroscopy of flame-arc, arc and spark sources at ICL whilst a research student under Prof A Fowler. Catal\'an determined wavelengths for 169 {\mn} lines in the spectral range 257 - 1760~nm (5677-38815 {\cm}), using these lines to identify 32 low-lying energy levels. Catal\'an was the first to observe and comment on the appearance of repeated patterns of transitions in the spectrum of manganese, which he called ``multiplets", which led to the general recognition of the multiplet structure of other complex spectra. In 1926, absorption measurements in the region 187 - 1761~nm  (5679 - 53342 {\cm}) by \citet{McLennon1926} extended the work of Catal\'an, increasing the number of known energy levels to 90 and the number of identified lines to 257. 


In 1945 Charlotte Moore produced a ``Multiplet Table of Astrophysical Interest" \citep{Moore1945} as part of her comprehensive review of atomic energy levels. Moore's comments on the lack of available data for Mn I prompted Catal\'an to set his student Olga Garcia-Riquelme the task of reassessing the data on Mn I.  In 1949, \citet{Garcia1949} compiled all known wavelengths, line intensities and energy level values of {\mn} from 15 different authors and analysed these to find 58 new energy levels. Fifteen years later, \citet{Catalan1964} published the last major work (note that this paper was finished after Catal\'an's death in 1957) on {\mn} wavelengths and energy levels. They reviewed the quality of all available {\mn} data (including unpublished work by T. Dunham Jr.) and re-analysed the data to identify further {\mn} transitions, bringing the total number of line wavelengths  to  2030 across the wavelength range 178 - 1761~nm (5677 - 56011 {\cm})  and resulting in determination of 404 energy levels in total.

The most recent published compilation of {\mn} energy levels is the 1985 work by \citet{Sugar1985a}, referred to from now on as S\&C. The majority of level values in the S\&C compilation are taken directly from \citet{Catalan1964}. As no energy level uncertainties were given by Catal\'an et al, S\&C estimated these from differences between Ritz and observed wavelengths to be $\pm$0.1 {\cm}. S\&C also included a small number of high lying $3d^64s(^7S)np$ energy levels, identified from low-resolution absorption spectra recorded by \citet{Ginter1978}, estimating their uncertainty to be $\pm$0.05 {\cm}. A more recent publication by \citet{Taklif1990} identified 53 new {\mn} lines in the IR between 0.82 and 1.97 $\mu$m (5068 - 11929 {\cm}), but using low-resolution grating spectra (resolution 0.008 nm), and no analysis to determine new energy levels was performed. 

Although the range of the previously existing spectral data for {\mn} extends from the vacuum UV (VUV) to the IR, the quality of the measurements varies immensely. Only approximately 5\% of the spectral lines in the visible have been measured with wavenumber uncertainty of $\leq$0.05 {\cm} or better. Indeed, the majority of lines are only known to an uncertainty of 0.1 {\cm} or worse. The accuracies of these data now fall far short of the requirements of high-resolution astrophysical spectral analyses. Furthermore, with the exception of a small number of lines measured to obtain HFS splitting factors (e.g. \citet{RBWMnHFS2005} and selected references therein), the majority of spectra in the literature have been measured at low resolutions, which are insufficient to resolve many of the complex features of the {\mn} spectrum. With high resolution FT spectroscopy it is now possible to reduce the uncertainties of transition wavelengths and energy levels of {\mn} by at least an order of magnitude.

This work addresses the critical need for {\mn} atomic data. As a result of an extensive term analysis of {\mn}, this paper presents new and improved energy level values and transition wavelengths for \mn, which will be invaluable in the interpretation of stellar and laboratory spectra across a wide range of astrophysical and applied contexts.

\section{Experimental details}
\label{sect:exp_details}
The FT and grating spectra used in this work were previously employed in the energy level analysis of {\mnii} by \citet{Liggins2021},  where full details of the spectral acquisition and calibration procedures are described in its Section 2. We provide a brief summary of these details in the following sections. These spectra provide a rich and well-characterised dataset that forms the foundation of the present term analysis of {\mn}. For completeness, a brief summary of the relevant spectra and the resulting linelists is given in the following sections.

In addition, to support the identification and confirmation of several energy levels in {\mn}, we incorporated the historical line list of \citet{Catalan1964}. Due to the lower accuracy of this legacy dataset compared to the new FTS and grating data, Calat\'an's lines were used primarily to check level assignments where only a small number of FTS or grating lines were available. All energy levels given in this paper have at least one FTS or newly-measured grating line. A discussion of our uncertainty estimates for Catal\'an's lines is provided in Section \ref{sect:catalan_unc}.

\subsection{Fourier Transform (FT) Spectroscopy}
\label{sect:fts}

High-resolution FT spectra were recorded at ICL and NIST, covering a broad spectral range from the IR to the VUV. Measurements in the visible to VUV region (24,005–58,990 {\cm}) were obtained using the ICL UV- and VUV-FT spectrometers \citep{Thorne1987, Thorne1996}, while IR spectra (1,820–26,632 {\cm}) were recorded at NIST using the 2m FT spectrometer \citep{Nave1997}. Additional lines were extracted from high-current FT spectra recorded by \citet{Kling_2000}, originally intended for measurements of {\mnii} transition probabilities. While self-absorption affected stronger lines in these high-current spectra, weaker {\mn} lines were present that were not observed in other spectra.  A summary of the FTS spectra used in this work is given in Table \ref{tab:keff1}.  

\begin{deluxetable*}{lllllllllll}  
\tablecaption{FT spectra used in the \mn\ analysis}
\tablewidth{\textwidth}
\tabletypesize{\scriptsize}
\tablehead{
\colhead{Date/} & \colhead{Region used} & \colhead{Instrument} & \colhead{Filter} & \colhead{Detector} &\colhead{Resolution} & \colhead{Current} & \colhead{Pressure} & \colhead{Gas}  & \colhead{$k_{\text{eff}}$~*}  & \colhead{Note $\dagger$} \\
\colhead{Serial no.} & \colhead{(cm$^{-1}$)} & \colhead{} & \colhead{} &\colhead{} &\colhead{(cm$^{-1}$)} & \colhead{(A)}  & \colhead{(Pa)} & \colhead{} & \colhead{($10^{-7})$}& \colhead{}
}                                                                                          
\startdata                                             
2001 Jan  25 no. 2  & 1820 - 9550       & NIST 2m    &     -  	  &  InSb  &0.01  & 1.2  & 75  & Ar &  3.2$^*$$\pm$ 0.2   & maah \\ 
2001 Jan  26 no. 3  & 1820 - 9550       & NIST 2m    &     -  	  &  InSb  &0.01  & 1.5  & 250 & Ne &  2.23$^*$$\pm$ 0.17   & manh \\ 
2001 Jan  19 no. 4  & 4010 - 17\,490    & NIST 2m    &     -  	  &  InSb  &0.009 & 1.5  & 250 & Ne &  2.50$^*$$\pm$ 0.16   & mcnh \\  
2001 Jan  22 no. 14 & 8520 - 24\,985    & NIST 2m    &     -  	  &  Si    &0.013 & 1.2  & 75  & Ar &  2.55$^*$$\pm$ 0.12   & meah \\ 
2001 Jan  16 no. 3  & 8443 - 25\,291    & NIST 2m    &     -  	  &  Si    &0.013 & 1.7  & 250 & Ne &  2.31$^*$$\pm$ 0.07   & menh \\ 
2001 Jan  16 no. 8  & 15\,918 - 24\,985 & NIST 2m    & CuSO$_{4}$ &  Si    &0.02  & 1.0  & 80  & Ar &  2.08$\pm$ 0.02   & mgah \\ 
2001 Jan  12 no. 16 & 15\,993 - 26\,632 & NIST 2m    & CuSO$_{4}$ &  Si    &0.02  & 2.0  & 250 & Ne &  3.41$\pm$ 0.04   & mgnh \\ 
2001 Mar  30        & 24\,007 - 29\,700 & ICL UV      &   Glass    & 1P28   &0.04  & 0.45 & 90  & Ar &  2.26$\pm$ 0.017  & mjah \\ 
2001 Mar  27        & 24\,057 - 29\,786 & ICL UV      &   Glass    & 1P28   &0.04  & 0.45 & 340 & Ne &  2.71$\pm$ 0.05   & mjnh \\ 
2001 Mar  27        & 24\,779 - 27\,567 & ICL UV      &   Glass    & 1P28   &0.04  & 0.20 & 340 & Ne &  2.47$\pm$ 0.06   & mjnl \\ 
2001 Apr  02        & 25\,421 - 41\,943 & ICL UV      &   UG5      & 1P28   &0.04  & 0.45 & 90  & Ar &  1.80$\pm$ 0.06   & mkah \\ 
2001 Mar  22        & 25\,355 - 41\,970 & ICL UV      &   UG5      & 1P28   &0.04  & 0.45 & 340 & Ne &  2.41$\pm$ 0.05   & mknh \\ 
2001 Mar  26        & 35\,690 - 35\,770 & ICL UV      &   UG5      & 1P28   &0.04  & 0.20 & 340 & Ne &  2.76$\pm$ 0.07   & mknl \\ 
2001 Apr  02        & 33\,897 - 49\,984 & ICL UV      &     -	  & R166   &0.04  & 0.45 & 90  & Ar &  2.21$\pm$ 0.12   & mlah \\ 
2001 Mar  21 no. 2  & 33\,897 - 52\,425 & ICL UV      &     -	  & R166   &0.04  & 0.45 & 340 & Ne &  2.21$\pm$ 0.16   & mlnh \\ 
2001 Mar  21 no. 1  & 35\,770 - 35\,806 & ICL UV      &     -	  & R166   &0.04  & 0.45 & 340 & Ne &  2.26$\pm$ 0.21   & mlnl \\ 
2001 Dec  04        & 33\,895 - 58\,990 & ICL VUV     &   BP185    & R1220  &0.05  & 0.45 & 340 & Ne &  0.81$\pm$ 0.29   & mmnh \\ 
2012 Sep  13        & 33\,897 - 58\,990 & ICL VUV     &   BP185    & R1220  &0.08  & 0.70 & 300 & Ne & 13.30$\pm$ 0.50   & 	-  \\ 
1998 Nov  12 no. 6  & 21\,480 - 37\,910 & NIST FT700 &     - 	  & R106UH &0.06  & 2.00 & 133 & Ne & -1.32$\pm$ 0.10   & 	-  \\ 
1998 Nov  12 no. 2  & 36\,800 - 47\,510 & NIST FT700 &     - 	  & R7154  &0.07  & 2.00 & 150 & Ne &  8.76$\pm$ 0.10   & 	-  \\ 
 \enddata
\tablecomments{$^{*}$$k_{\text{eff}}$, calibration correction factor for the spectrum with its uncertainty, where superscript * indicates a newly revised  $k_{\text{eff}}$, detailed in Section \ref{sect:IR_recalib}. The uncertainty in $k_{\text{eff}}$ is the propagated uncertainty through the calibration chain. $^{\dagger}$Name of spectrum as given in \citet{RBW2003}.
}
\label{tab:keff1}
\end{deluxetable*}                               

Water-cooled hollow cathode lamps (HCLs) with argon or neon carrier gases served as the emission sources for the spectra. To mitigate the brittleness of pure Mn, cathodes were fabricated from Mn–Cu (95:5) and Mn–Ni (88:12) alloys for the NIST and ICL spectra, respectively. The HCLs produced emission from {\mn}, {\mnii}, neutral and singly ionised carrier gas lines (Ar or Ne), and impurity lines from the cathode alloy metals. Optimal operating conditions were 0.9 mbar of Ar or 3.4 mbar of Ne at 450 mA. Additional low-current runs at 200 mA, under the same pressures, were performed to assess self-absorption in strong metal lines. The resolution of the spectra ranged from 0.009 {\cm} in the IR to 0.08 {\cm} in the VUV.

Intensity calibration was achieved by recording the spectrum of a radiometrically calibrated standard lamp immediately before and after each HCL measurement. Comparison of the averaged standard lamp spectrum to the calibrated radiance of the standard lamp enabled determination of the average instrument response function over the measurement period. Relative line intensities were obtained by dividing observed intensities by the response function. A tungsten standard lamp calibrated by the National Physical Laboratory (NPL), UK, was used at ICL for calibration of spectra in the IR to visible regions. Another tungsten lamp, calibrated by Optronics Laboratories Inc., USA, was used for all the NIST spectra. Deuterium standard lamps with either silica or MgF$_2$ windows were used in the visible to UV and VUV regions at ICL, with calibrations performed by NPL and the Physikalisch-Technische Bundesanstalt (PTB), Germany, respectively.

All spectra were placed on a common relative intensity scale using strong metal lines in overlapping regions to determine scaling factors. The relative populations of energy levels and resulting line intensities depend on the carrier gas, current, and pressure of each HCL. The relative line intensities presented here are approximate and should not be used for precise calculations of level branching fractions or transition probabilities.

The wavenumber, integrated intensity, full width at half maximum (FWHM), and signal-to-noise ratio (S/N) were measured for each spectral line using the spectral analysis software Xgremlin \citep{Nave2015}. For lines without observed asymmetries, least-squares fits to Voigt profiles were performed. For asymmetric lines and those with significant HFS, centre-of-gravity (CoG) fits were used to determine the centroid wavenumber. CoG fits were used for the majority of lines in the FT spectra. 

For certain critical transitions, such as resonance lines, key intercombination lines, or cases where line identifications were uncertain, a detailed HFS fit was required, especially when the measured wavenumber uncertainty was much higher than expected from the signal-to-noise ratio. In these instances, we employed the HFS fitting routines in Xgremlin, originally developed by \citet{Pulliam1979}. These routines perform iterative fits using up to eight adjustable parameters: the magnetic dipole (A) and electric quadrupole (B) constants for both upper and lower levels, the central wavenumber, full width at half maximum (FWHM), Voigt profile damping factor, and signal-to-noise ratio. Parameters could be fixed to known values if available, and the software allowed simultaneous fitting of up to three overlapping transitions per spectral feature. The fitting process provided uncertainties for all derived parameters.

Wavenumber calibration was performed using  26 \ion{Ar}{2} reference lines in the visible region \citep{Learner1988}, based on measurements by \citet{Whaling1995}. Calibrated Mn lines from Mn–Ar spectra were then used to transfer the calibration to Mn–Ne spectra, extending coverage into the IR and VUV. Each step in this cross-calibration process introduces a cumulative uncertainty in the wavenumber positions. The total calibration uncertainty of a spectral line is the combination of the error in the applied calibration correction for each spectrum and the uncertainties of the initial reference wavenumbers, which trace back to the standards reported by \citet{Whaling1995} with an accuracy of approximately one part in $10^8$. The total wavenumber uncertainty of each spectral line is then the combination in quadrature of the calibration uncertainty of the line's spectrum with the statistical uncertainty of the line fit.

In total, 18,948 spectral lines were measured across all FT spectra. Multiple observations of the same line were merged using a weighted average, with weights inversely proportional to the signal-to-noise ratio. This yielded 10,432 unique lines in the final linelist, encompassing all species present in the HCL plasmas: {\mn}, {\mnii}, carrier gases, cathode alloy constituents, and impurities.

\subsubsection{Recalibration of IR FTS Spectra}
\label{sect:IR_recalib}
It was noticed during the initial energy level optimisation that the values of the $3d^5(^6S)4s4p(^3P)\;\,z^6P$ levels were inconsistent, depending on whether they were determined via the ground level, $3d^54s^2\;\,a^6S_{\nicefrac{5}{2}}$, with lines around 400~nm, or via the $(^5D)4s\;\, a^6D$ term with lines in the IR around 1350~nm. Accurate values for these levels not only set the foundation for the entire term system, but also determine improved wavelengths for parity-forbidden lines between the $a^6S$ and $a^6D$ terms.

Whilst there are various reasons why the lines from the $z^6P$ term may be inconsistent, the large separation of the lines to the $a^6S$ and $a^6D$ terms led us to re-examine the wavenumber calibration of the IR spectra. Wavenumber calibration constants, $k_{\text{eff}}$, are calculated from observed wavenumbers, $\sigma_{\text{obs}}$, and reference wavenumbers, $\sigma_{\text{ref}}$, where
\begin{equation}
    k_{\text{eff}} = 1- \sigma_{\text{obs}}/\sigma_{\text{ref}}
\end{equation}
These values are given in Table 3.2 of \citet{Liggins_2017} (hereafter L17), and in Table 1 of \citet{Liggins2021}. Corresponding plots of the value of $k_{\text{eff}}$ against wavenumber are given in Figs. 3.3 to 3.18 of L17. The primary wavenumber standards, given in Table 3.1 of L17, were 26 lines of \ion{Ar}{2} from \citet{Whaling1995}. Three spectra (mjah, mgah, and meah) were calibrated directly from these standards and the calibration plots are shown in Figs 3.3, 3.4, and 3.5 of L17 respectively. The remaining spectra were calibrated using strong, unblended, Mn lines found in these spectra, and the calibration chain is given in Table 3.5 of L17 and Table 1 of \citet{Liggins2021}. This table shows that the three spectra in the IR (manh, maah, and mcnh) were all calibrated from the spectrum meah, whereas all the other spectra in Table 3.2 of L17 were calibrated via mjah, either directly or using overlapping spectra. 

\begin{figure*}
    \centering
    \includegraphics[width=0.45\textwidth]{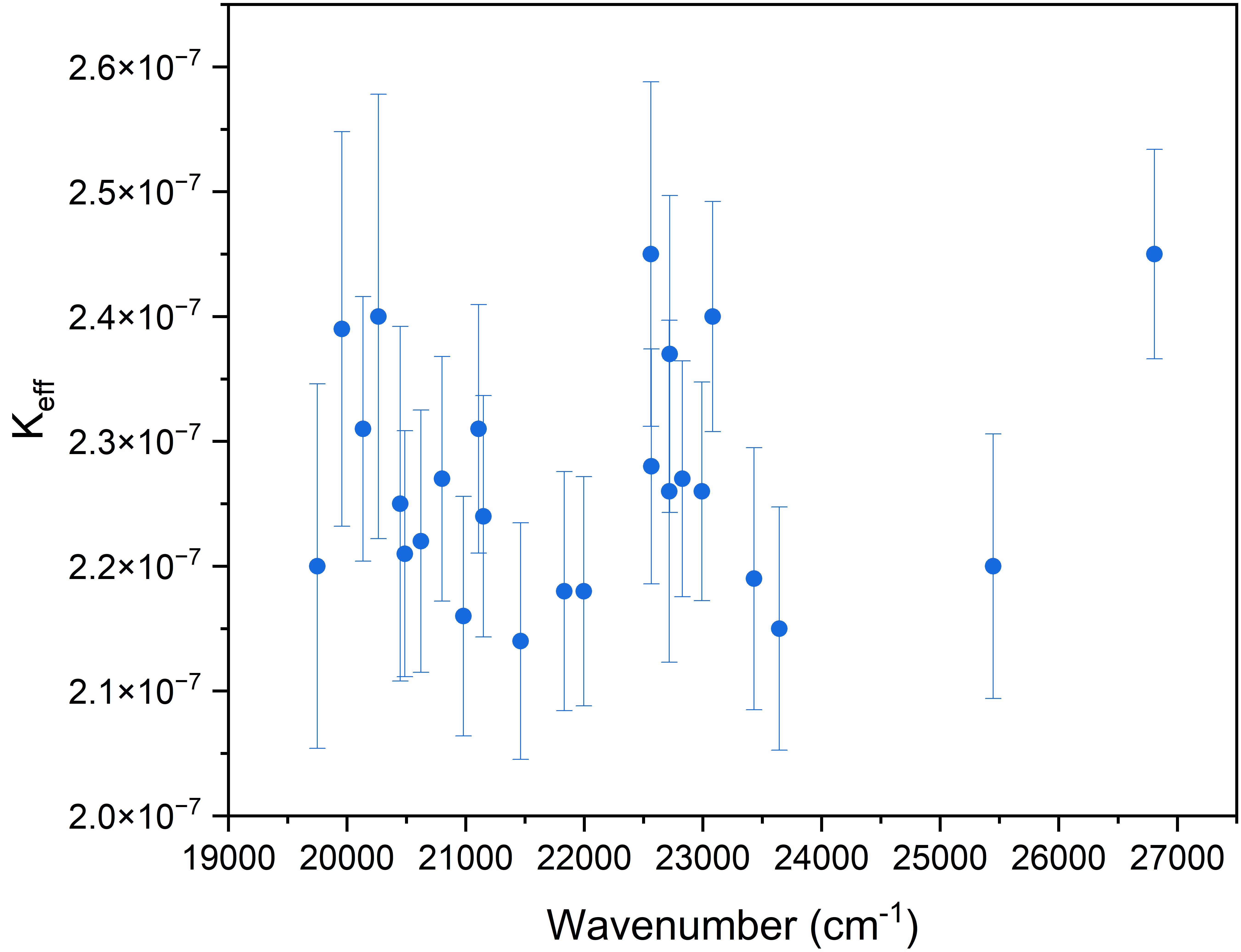}
    \hspace{0.03\textwidth}
    \includegraphics[width=0.465\textwidth]{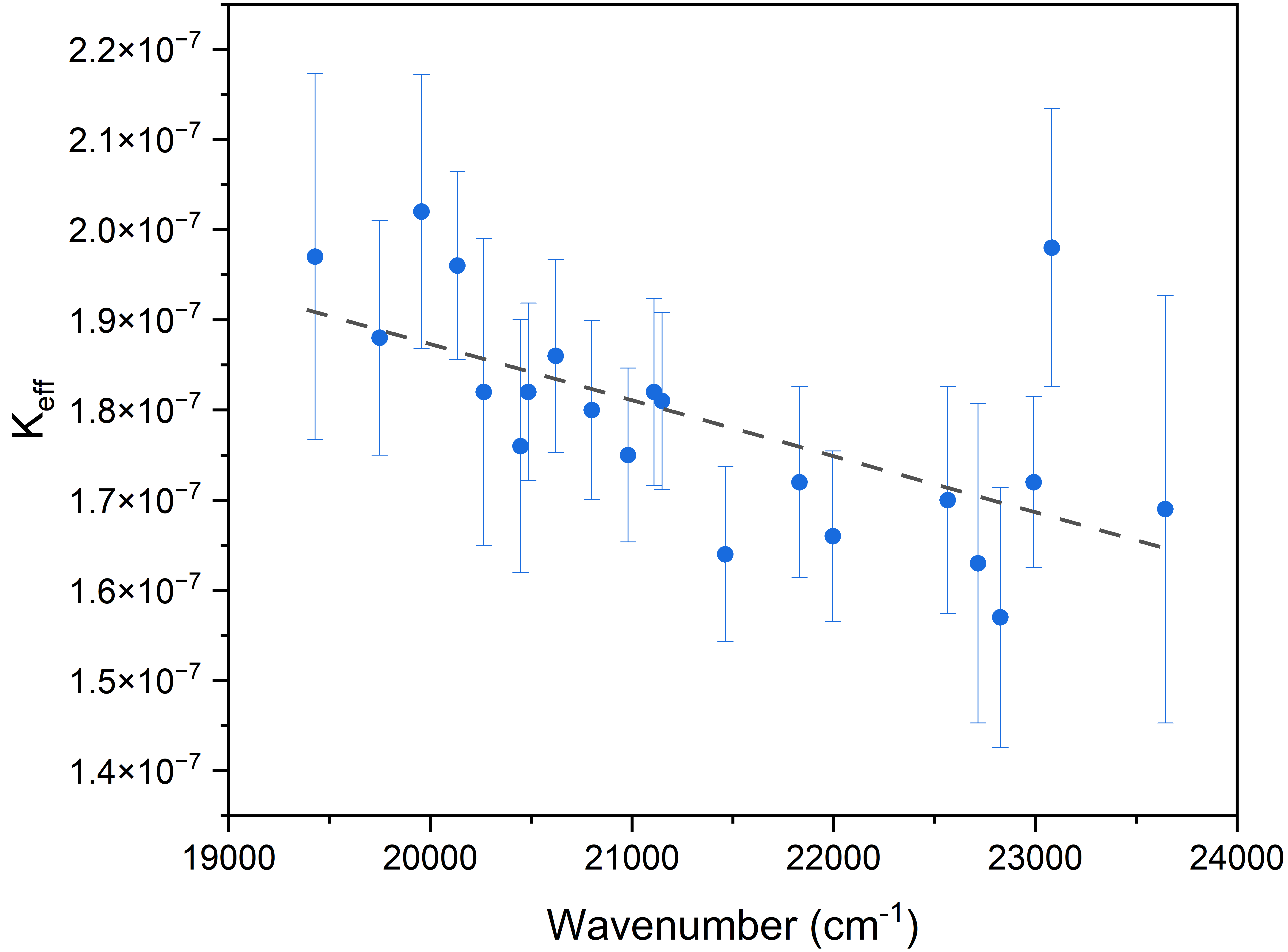}
    \caption{Original calibration of spectra mjah (left) and meah (right) using \ion{Ar}{2} lines from \citet{Liggins_2017}. The dotted line in the meah plot shows a linear fit to the $k_{\text{eff}}$ data with a slope of $-6.2\times10^{-12}$ per {\cm}.}
    \label{fig:menh_orig}
\end{figure*}

The calibration plots of mjah and meah are given in Figs 3.3 and 3.4 of L17 and are reproduced here in Fig. \ref{fig:menh_orig}. It can be seen that while the value of $k_{\text{eff}}$ is independent of wavenumber for spectrum mjah, the plot for spectrum meah shows a small slope of about $-6.2\times10^{-12}\sigma$ from 19~000~{\cm} to 24~000~{\cm}. This slope would result in an error of 0.0007~{\cm} at 7300~{\cm}, roughly three times the uncertainty of strong lines in this region. The wavelength-dependent nature of the deviation suggests that the issue with spectrum meah is more likely due to illumination shifts, rather than plasma effects, which would manifest as energy level dependent variations.

\begin{figure}
    \centering
    \includegraphics[width=0.95\linewidth]{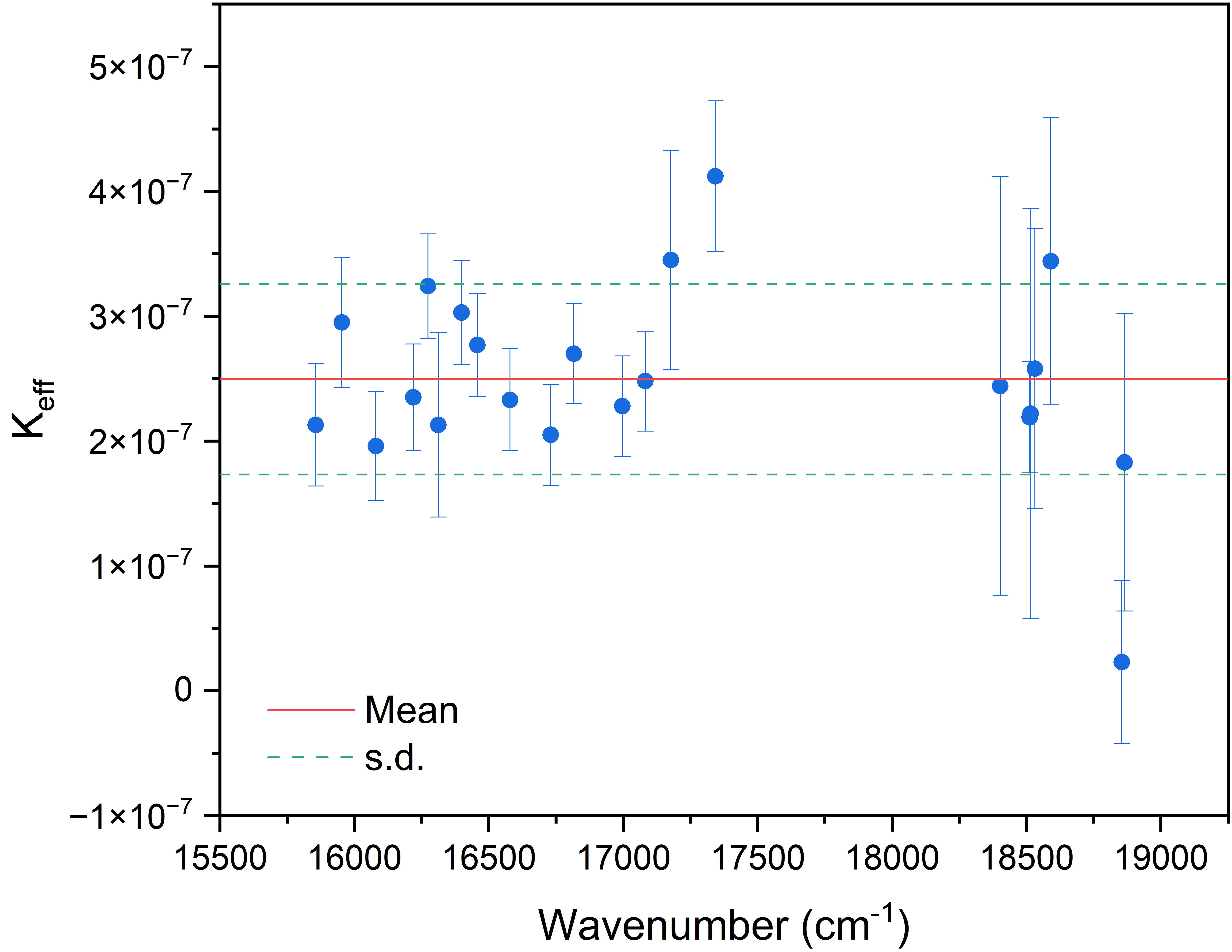}
    \caption{Recalibration of spectrum mcnh from mjnh for lines with S/N $> 50$.}
    \label{fig:menh_recal}
\end{figure}

Fortunately, there are sufficient overlapping lines between spectrum mjnh, calibrated directly from mjah, and mcnh to provide a revised calibration for this spectrum that omits the problematic spectrum meah. This calibration is shown in Fig. \ref{fig:menh_recal}, and has an offset of about 8 parts in $10^8$ with respect to the value given in Table 3.2 of L17. The recalibrated spectrum mcnh was used to adjust the calibration of manh and maah in the IR region, and the calibration of meah and menh between 7000~{\cm} to 18~000~{\cm}.  Details of the revised values of $k_{\text{eff}}$ are given in Table \ref{tab:keffcorrdat}. The values of $k_{\text{eff}}$ for all other spectra are unchanged from Table 3.2 of L17 and Table 1 of \citet{Liggins2021}, but the details of all spectra used in this work are included in Table \ref{tab:keff1} here for completeness. 

\begin{deluxetable*}{llllrrrrr}  
\tablecaption{Revised values of $k_{\text{eff}}$ for IR FT spectra}
\tablewidth{450pt}
\tablehead{
 \colhead{Spectrum} & \colhead{Calibrated from} & \colhead{Old $k_{\text{eff}}$ } & \colhead{New $k_{\text{eff}} $} & \colhead {S.D. } &\colhead{S.E.} & \colhead{S.E.$_{prop}$ } & \colhead{N$_{fit}$} & \colhead{S/N$_{min}$} \\
\colhead{} &  \colhead{} & \colhead{ (10$^{-7}$) } & \colhead{ ($10^{-7}$)} & \colhead {( 10$^{-8}$ )} &\colhead{(10$^{-8}$)} & \colhead{(10$^{-8}$)} & \colhead{} & \colhead{} 
}              
\startdata                                                   
mcnh  &  mjnh & 1.68(5)      & 2.50(16)    & 7.62  & 1.62   & 2.1 & 22  & 50   \\
manh  &  mcnh & 1.31(11)     & 2.23(17)    & 7.01  & 1.65   & 3.8 & 18  & 50   \\
maah  &  mcnh & 2.50(9)      & 3.2(2)      & 9.71  & 2.23   & 4.1 & 19  & 50   \\
menh  &  mcnh & 1.49(4)      & 2.31(7)     & 9.25  & 0.69   & 2.8 & 180 & 50   \\
meah  &  mcnh & 1.78(3)      & 2.55(12)    & 7.95  & 1.18   & 3.3 & 45  & 50   \\
 \enddata
\tablecomments{The old and newly revised $k_{\text{eff}}$, calibration correction factors, for the listed IR spectra, with their standard deviation (S.D.) and standard error (S.E.). The final propagated standard error (S.E.$_{prop}$) is the propagated uncertainty through the calibration chain by linear addition of calibration uncertainty to that of the previous spectrum. $N_{fit}$ is the number of lines used to calculate the calibration factor and S/N$_{min}$ gives the minimum S/N of the lines used in the calculation of $k_{\text{eff}}$.}
\label{tab:keffcorrdat}
\end{deluxetable*}                               

It should be noted that very few lines of {\mnii} are affected by this change as lines of {\mnii} are typically weaker than those of {\mn} in the IR region and thus have statistical uncertainties larger than the calibration correction. Table \ref{tab:mn2_recalib_lines} gives the lines of {\mnii} that have changed from \citet{Liggins2021} by more than one standard uncertainty following the recalibration of spectra described above.

\begin{table*}[h]
    \centering
    \caption{Classified {\mnii} line wavenumbers affected by the recalibration of IR FT spectra}
    \begin{tabular}{lllll}
        \hline\hline
        Previous Wavenumber & Recalibrated Wavenumber & \multicolumn{2}{c}{Transition} & Spectrum \\ \cline{3-4}
         ({\cm})& ({\cm}) & Lower & Upper &   \\ \hline
6344.9269(5) & 6344.9274(5) & $3d^5(^6S)4d\;\;e^7D_{\nicefrac{3}{2}}$ & $3d^5(^6S)5p\;\;x^7P\degree_{\nicefrac{2}{2}}$ & mcnh \\
6401.9860(4) & 6401.9865(4) & $3d^5(^6S)4d\;\;e^7D_{\nicefrac{4}{2}}$ & $3d^5(^6S)5p\;\;x^7P\degree_{\nicefrac{3}{2}}$ & mcnh \\
6488.2264(4) & 6488.2269(4) & $3d^5(^6S)4d\;\;e^7D_{\nicefrac{5}{2}}$ & $3d^5(^6S)5p\;\;x^7P\degree_{\nicefrac{4}{2}}$ & mcnh \\
7745.6913(4) & 7745.6919(4) & $3d^5(^6S)5d\;\;^5D_{\nicefrac{2}{2}}$ & $3d^5(^6S)5f\;\;^5F\degree_{\nicefrac{3}{2}}$ & mcnh \\
7752.6648(4) & 7752.6654(4) & $3d^5(^6S)5d\;\;^5D_{\nicefrac{3}{2}}$ & $3d^5(^6S)5f\;\;^5F\degree_{\nicefrac{4}{2}}$ & mcnh \\
7760.5387(4) & 7760.5393(4) & $3d^5(^6S)5d\;\;^5D_{\nicefrac{4}{2}}$ & $3d^5(^6S)5f\;\;^5F\degree_{\nicefrac{5}{2}}$ & mcnh \\
8505.6310(4) & 8505.6317(4) & $3d^5(^6S)5d\;\;^7D_{\nicefrac{5}{2}}$ & $3d^5(^6S)5f\;\;^7F\degree_{\nicefrac{6}{2}}$ & mcnh \\
8510.8928(4) & 8510.8935(4) & $3d^5(^6S)5d\;\;^7D_{\nicefrac{4}{2}}$ & $3d^5(^6S)5f\;\;^7F\degree_{\nicefrac{5}{2}}$ & mcnh \\
10090.7472(6) & 10090.7480(6) & $3d^5(^6S)4f\;\;^5F\degree_{\nicefrac{5}{2}}$ & $3d^5(^6S)5g\;\;^5G_{\nicefrac{6}{2}}$ & mcnh \\
10131.9471(6) & 10131.9479(6) & $3d^5(^6S)4f\;\;^7F\degree_{\nicefrac{1}{2}}$ & $3d^5(^6S)5g\;\;^7G_{\nicefrac{1}{2}}$ & mcnh \\
10132.0498(6) & 10132.0506(6) & $3d^5(^6S)4f\;\;^7F\degree_{\nicefrac{2}{2}}$ & $3d^5(^6S)5g\;\;^7G_{\nicefrac{1}{2}}$ & mcnh \\
10132.1628(6) & 10132.1636(6) & $3d^5(^6S)4f\;\;^7F\degree_{\nicefrac{1}{2}}$ & $3d^5(^6S)5g\;\;^7G_{\nicefrac{2}{2}}$ & mcnh \\
10132.2471(7) & 10132.2479(7) & $3d^5(^6S)4f\;\;^7F\degree_{\nicefrac{2}{2}}$ & $3d^5(^6S)5g\;\;^7G_{\nicefrac{3}{2}}$ & mcnh \\
10132.3160(5) & 10132.3168(5) & $3d^5(^6S)4f\;\;^7F\degree_{\nicefrac{3}{2}}$ & $3d^5(^6S)5g\;\;^7G_{\nicefrac{4}{2}}$ & mcnh \\
10132.3777(5) & 10132.3785(5) & $3d^5(^6S)4f\;\;^7F\degree_{\nicefrac{5}{2}}$ & $3d^5(^6S)5g\;\;^7G_{\nicefrac{6}{2}}$ & mcnh \\
10132.4341(5) & 10132.4349(5) & $3d^5(^6S)4f\;\;^7F\degree_{\nicefrac{4}{2}}$ & $3d^5(^6S)5g\;\;^7G_{\nicefrac{5}{2}}$ & mcnh \\
10132.4341(5) & 10132.4349(5) & $3d^5(^6S)4f\;\;^7F\degree_{\nicefrac{6}{2}}$ & $3d^5(^6S)5g\;\;^7G_{\nicefrac{7}{2}}$ & mcnh \\
10523.0477(5) & 10523.0486(5) & $3d^5(^6S)5s\;\;e^5S_{\nicefrac{2}{2}}$ & $3d^5(^6S)5p\;\;w^5P\degree_{\nicefrac{3}{2}}$ & menh \\
10562.1796(5) & 10562.1805(5) & $3d^5(^6S)5s\;\;e^5S_{\nicefrac{2}{2}}$ & $3d^5(^6S)5p\;\;w^5P\degree_{\nicefrac{2}{2}}$ & menh \\
10586.3327(5) & 10586.3336(5) & $3d^5(^6S)5s\;\;e^5S_{\nicefrac{2}{2}}$ & $3d^5(^6S)5p\;\;w^5P\degree_{\nicefrac{1}{2}}$ & menh \\
11335.3143(6) & 11335.3152(6) & $3d^5(^6S)5s\;\;e^7S_{\nicefrac{3}{2}}$ & $3d^5(^6S)5p\;\;x^7P\degree_{\nicefrac{2}{2}}$ & mcnh \\
11400.4485(6) & 11400.4494(6) & $3d^5(^6S)5s\;\;e^7S_{\nicefrac{3}{2}}$ & $3d^5(^6S)5p\;\;x^7P\degree_{\nicefrac{3}{2}}$ & mcnh \\
11497.4267(6) & 11497.4276(6) & $3d^5(^6S)5s\;\;e^7S_{\nicefrac{3}{2}}$ & $3d^5(^6S)5p\;\;x^7P\degree_{\nicefrac{4}{2}}$ & mcnh \\
13751.8292(7) & 13751.8303(7) & $3d^5(^6S)5p\;\;w^5P\degree_{\nicefrac{2}{2}}$ & $3d^5(^6S)5d\;\;^5D_{\nicefrac{3}{2}}$ & mcnh \\
13784.7483(7) & 13784.7494(7) & $3d^5(^6S)5p\;\;w^5P\degree_{\nicefrac{3}{2}}$ & $3d^5(^6S)5d\;\;^5D_{\nicefrac{4}{2}}$ & mcnh \\
13790.9586(11) & 13790.9597(11) & $3d^5(^6S)5p\;\;w^5P\degree_{\nicefrac{3}{2}}$ & $3d^5(^6S)5d\;\;^5D_{\nicefrac{3}{2}}$ & mcnh \\
13841.5649(8) & 13841.5660(8) & $3d^5(^6S)5p\;\;x^7P\degree_{\nicefrac{4}{2}}$ & $3d^5(^6S)5d\;\;^7D_{\nicefrac{4}{2}}$ & mcnh \\
13846.7023(7) & 13846.7034(7) & $3d^5(^6S)5p\;\;x^7P\degree_{\nicefrac{4}{2}}$ & $3d^5(^6S)5d\;\;^7D_{\nicefrac{5}{2}}$ & mcnh \\
13931.8750(11) & 13931.8761(11) & $3d^5(^6S)5p\;\;x^7P\degree_{\nicefrac{3}{2}}$ & $3d^5(^6S)5d\;\;^7D_{\nicefrac{2}{2}}$ & mcnh \\
13938.5424(7) & 13938.5435(7) & $3d^5(^6S)5p\;\;x^7P\degree_{\nicefrac{3}{2}}$ & $3d^5(^6S)5d\;\;^7D_{\nicefrac{4}{2}}$ & mcnh \\
13995.2053(7) & 13995.2064(7) & $3d^5(^6S)5p\;\;x^7P\degree_{\nicefrac{2}{2}}$ & $3d^5(^6S)5d\;\;^7D_{\nicefrac{1}{2}}$ & mcnh \\
13997.0068(7) & 13997.0079(7) & $3d^5(^6S)5p\;\;x^7P\degree_{\nicefrac{2}{2}}$ & $3d^5(^6S)5d\;\;^7D_{\nicefrac{2}{2}}$ & mcnh \\
13999.7948(7) & 13999.7959(7) & $3d^5(^6S)5p\;\;x^7P\degree_{\nicefrac{2}{2}}$ & $3d^5(^6S)5d\;\;^7D_{\nicefrac{3}{2}}$ & mcnh \\
16306.5871(8) & 16306.5884(8) & $3d^5(^6S)4d\;\;e^5D_{\nicefrac{1}{2}}$ & $3d^5(^6S)4f\;\;^5F\degree_{\nicefrac{2}{2}}$ & menh \\
16311.2760(13) & 16311.2773(13) & $3d^5(^6S)4d\;\;e^5D_{\nicefrac{2}{2}}$ & $3d^5(^6S)4f\;\;^5F\degree_{\nicefrac{2}{2}}$ & menh \\
16312.0712(8) & 16312.0725(8) & $3d^5(^6S)4d\;\;e^5D_{\nicefrac{2}{2}}$ & $3d^5(^6S)4f\;\;^5F\degree_{\nicefrac{3}{2}}$ & menh \\
16318.7747(10) & 16318.7760(10) & $3d^5(^6S)4d\;\;e^5D_{\nicefrac{3}{2}}$ & $3d^5(^6S)4f\;\;^5F\degree_{\nicefrac{3}{2}}$ & menh \\
16319.7153(8) & 16319.7166(8) & $3d^5(^6S)4d\;\;e^5D_{\nicefrac{3}{2}}$ & $3d^5(^6S)4f\;\;^5F\degree_{\nicefrac{4}{2}}$ & menh \\
16327.8551(13) & 16327.8564(13) & $3d^5(^6S)4d\;\;e^5D_{\nicefrac{4}{2}}$ & $3d^5(^6S)4f\;\;^5F\degree_{\nicefrac{4}{2}}$ & menh \\
16328.8579(8) & 16328.8592(8) & $3d^5(^6S)4d\;\;e^5D_{\nicefrac{4}{2}}$ & $3d^5(^6S)4f\;\;^5F\degree_{\nicefrac{5}{2}}$ & menh \\
16890.3253(9) & 16890.3267(9) & $3d^5(^6S)4p\;\;z^7P\degree_{\nicefrac{4}{2}}$ & $3d^44s^2\;\;c^5D_{\nicefrac{4}{2}}$ & mcnh, menh\\
&&&& mgnh, k1 \\
17289.8857(10) & 17289.8871(10) & $3d^5(^4F)4p\;\;x^5D\degree_{\nicefrac{4}{2}}$ & $3d^5(^6S)5d\;\;^7D_{\nicefrac{3}{2}}$ & mcnh, menh\\
&&&& mgnh, mgah \\
&&&& meah, k1 \\
\hline
    \end{tabular}
    \label{tab:mn2_recalib_lines}
\end{table*}

\subsection{Grating Spectroscopy}
\label{sect:grating}
Grating spectra of a high-current, water-cooled hollow cathode lamp (HCL) were recorded over the range 28,385–121,728 {\cm} (82 to 353 nm) using the 10.7~m normal-incidence vacuum spectrometer (NIVS) at NIST, USA. To suppress second-order diffraction from shorter wavelengths, MgF$_2$ or fused silica filters were used above 50,000 {\cm}. The HCL employed a Mn/Ni foil placed in a Cu cathode and was operated at a current of 2 A with 2 mbar of neon as the carrier gas. Spectra were recorded onto Kodak SWR photographic plates with 60-minute exposures. Full details are given in Table 2 of \citet{Liggins2021}.

Spectral line positions were measured using a Grant Instruments spectrum plate comparator equipped with a digital encoder. In addition to positional measurements, a visual estimate of line intensity was recorded for each feature. These intensity estimates are approximate and should be regarded only as qualitative indicators of relative line strength.

For wavelength calibration in the range 117–254 nm, spectra of a Pt–Ne HCL were recorded adjacent to the Mn exposures, allowing calibration via Pt standard lines \citep{Sansonetti1992}. For wavelengths below 117 nm, where no Pt–Ne spectra are available, calibration was performed using \ion{Cu}{2} Ritz wavelengths derived from \citet{atoms5010009}. For the remaining regions, {\mnii} Ritz wavelengths from \citet{Liggins2021} were used.

Uncertainties for each grating line were estimated based on line quality and photographic plate darkening. Lines with strengths between 5 and 60—sufficiently strong for accurate wavelength measurement but not saturated—were assigned an uncertainty of 0.0002 nm (equivalent to 0.05 {\cm} at 50,000 {\cm}). Lines outside this range, or those exhibiting visual issues such as haze, broadening, or blending, were assigned a larger uncertainty of 0.0005 nm (0.125 {\cm} at 50000 {\cm}).

In total, 10426 FTS + 11397 grating lines were included in the final linelist.

\subsection{Grating Data from \citet{Catalan1964}}
\label{sect:catalan_unc}

\begin{figure*}
    \centering
    \includegraphics[width=0.45\textwidth]{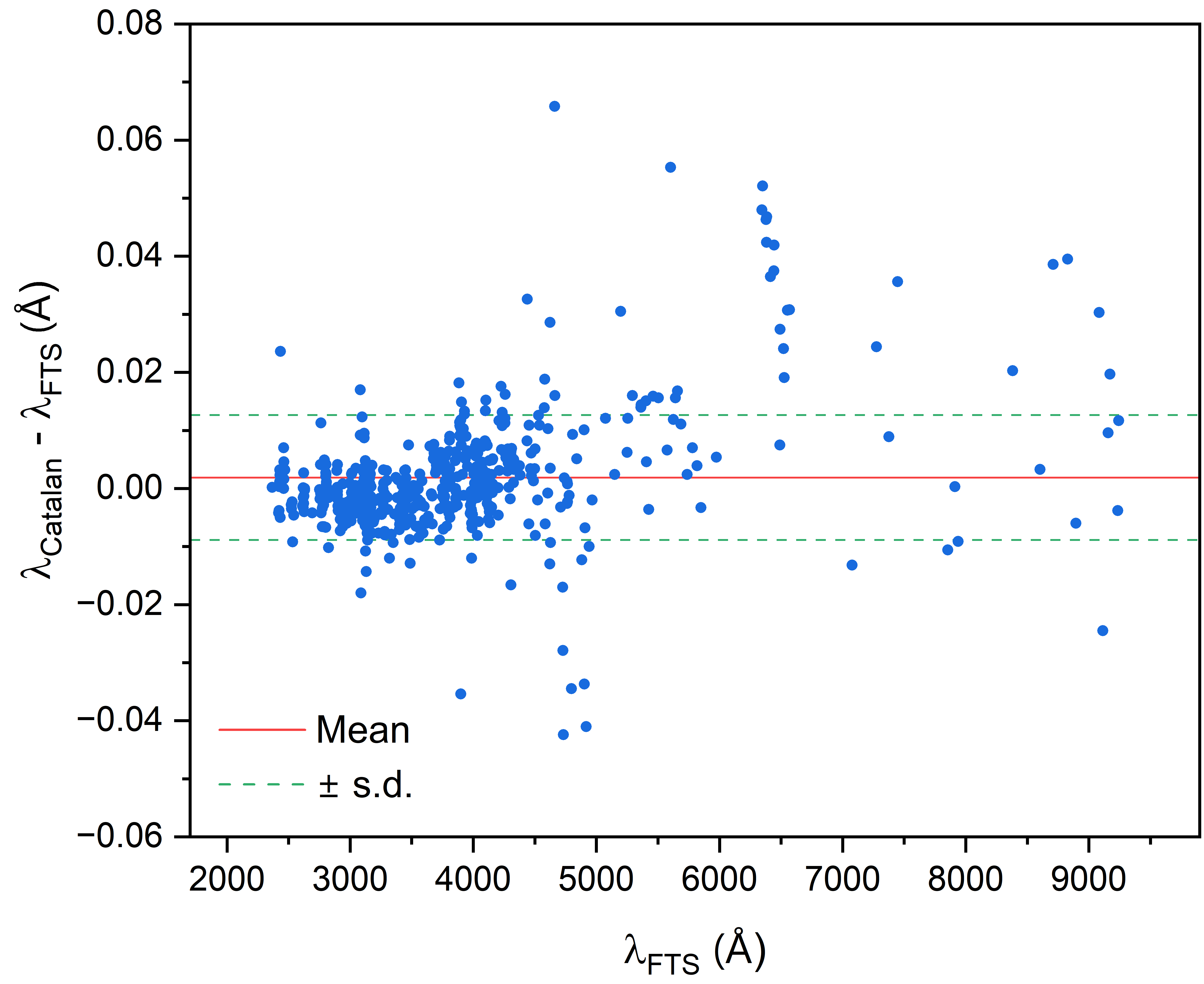}
    \hspace{0.03\textwidth}
    \includegraphics[width=0.465\textwidth]{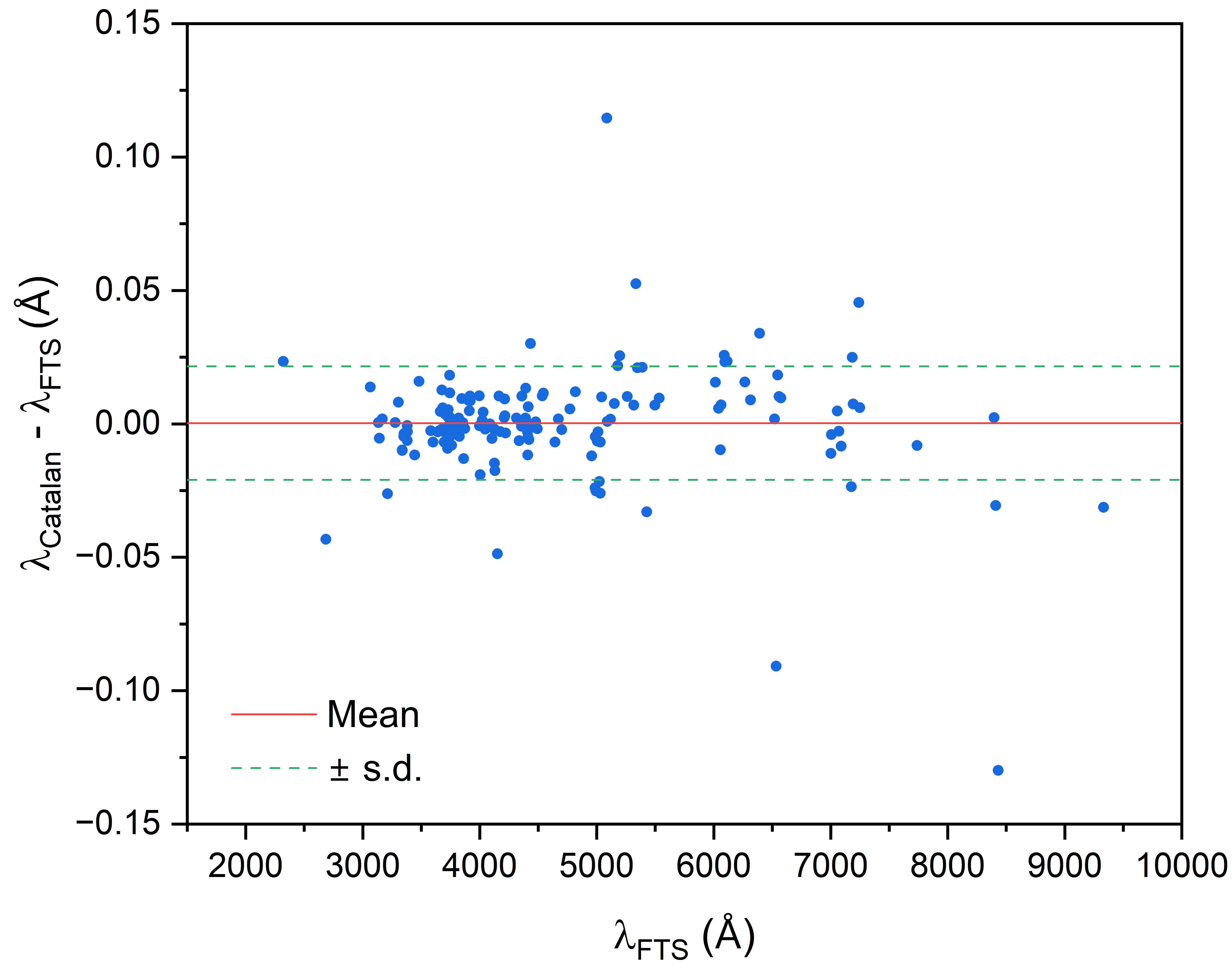}
    \caption{Comparisons between previously published  grating wavelengths \citep{Catalan1964} and FTS wavelengths \citep{Liggins2021}. Left: Lines with no identified issues. Right: Lines with issues (e.g. hazy, broadened).}
    \label{fig:cat_comp}
\end{figure*}

To supplement the new {\mn} FTS and grating measurements, selected transitions from the historical line list of \citet{Catalan1964} were added to our linelist. These transitions do not appear in any of the FTS or grating spectra described above, likely due to differences in source conditions or measurement sensitivity. To ensure consistency, the wavelength calibration of the {\mn} lines from \citet{Catalan1964} was carefully evaluated by matching common transitions to those observed in the FTS spectra. The distribution of wavelength differences was used to estimate wavelength uncertainties for Catal\'an's lines, as no uncertainties were reported in the original publication. Comparison plots between FTS wavelengths and Catal\'an's lines with no identified issues (given as line comments in \citet{Catalan1964}) and for lines \textit{with} identified issues are given in Figure~\ref{fig:cat_comp}. There is a small offset of the mean wavelength difference of $\sim$ 0.0018 {\AA} in both plots, but as this shift is an order of magnitude less than the spectral line uncertainties, we decided not to shift the wavelengths given in \citet{Catalan1964}.

Uncertainties for the Catal\'an lines were assigned based on qualitative assessments of line quality described in the original linelist. Lines flagged as problematic (e.g. hazy, broadened, or blended) were assigned an uncertainty of 0.003~nm, while all other lines were given an estimated uncertainty of 0.001~nm. In total, 718 {\mn} lines from \citet{Catalan1964} were not found in our grating or FTS spectra and were therefore included in our initial energy level analysis. Of these, 415 lines were used directly to support level identifications, and resolve ambiguities in term assignments. No energy levels in this work were determined using transitions observed exclusively in the Catal\'an spectra.

\subsection{Final Linelist}
The final linelist compiled for this work includes transitions from FT spectroscopy, grating spectroscopy, and selected grating historical measurements. In total, 24237 unique spectral lines were identified and characterised, spanning a broad wavenumber range, from 1820 {\cm} to 121717 {\cm} (821~nm to 54916~nm). Each line was assigned a wavenumber, relative intensity, S/N, FWHM, and a wavenumber uncertainty based on the measurement method and line quality. This comprehensive linelist forms the basis for the term analysis of {\mn} presented in this work.

\section{{\mn} Spectral Analysis}
\label{sect:analysis}
\subsection{Line Identification}
\label{sect:line_ident}

The first stage of the term analysis of {\mn} involved identifying transitions in the compiled linelist. Transitions were observed from neutral and singly ionised manganese ({\mn} and {\mnii}), as well as from carrier gases and impurity elements originating from the cathode alloys used in the HCLs.

Initial identification of {\mn} lines was carried out by comparing observed wavenumbers with Ritz values derived from the compiled energy levels of S\&C. {\mnii} transitions were identified using Ritz wavenumbers based on the extensive energy levels of \citet{Liggins2021}. Together, {\mn} and {\mnii} lines accounted for the vast majority of observed transitions.

Carrier gas lines were identified using published references: \ion{Ar}{1} and \ion{Ar}{2} from \citet{Whaling2002, Whaling1995}, \ion{Ne}{1} from \citet{Saloman2005}, and \ion{Ne}{2} from \citet{Kramida2006}. Impurity lines from cathode alloy constituents were identified using \citet{Litzen1993} for \ion{Ni}{1}, \citet{Clear_2022, Clear_2023} for \ion{Ni}{2}, \citet{NIST_ASD} for \ion{Cu}{1}, and \citet{atoms5010009} for \ion{Cu}{2}.

In the grating spectra above 120 nm, spectral lines of \ion{Ni}{2} and \ion{Cu}{2} were generally weak due to poor excitation of these species in neon under those conditions. However, strong \ion{Cu}{2} lines were observed below 120 nm, attributed to the high population of low-lying levels of \ion{Cu}{2} facilitated by the Ne carrier gas.

The complex term analysis process, described in sections \ref{sect:mn_structure}, \ref{sect:level_optimisation} and \ref{sect:new_levels}, is iterative, and results eventually in a final spectrum linelist of identified {\mn} lines.

\subsection{The {\mn} Spectrum}
\label{sect:mn_spectrum}

After the term analysis procedure, in total, 2186 spectral lines have been classified as {\mn} transitions. Of these, 1642 lines were measured using high-resolution FTS. A small number of lines have been multiply identified where a single assignment could not be conclusively made. 

All observed classified {\mn} transitions are listed in Table \ref{tab:linelist}. The first column indicates the method of spectral line measurement, with a key provided in the table footnote. The second column gives the integrated relative intensity of each line. For FTS lines, designated ``F", intensities were calibrated using standard lamp spectra. Grating lines recorded on photographic plates, designated ``G", have intensities based on visual estimates of plate darkening. Notes on line appearance (e.g. hazy, broadened, blended) are indicated by appended letters, with definitions provided in the footnote. For lines from \citet{Catalan1964}, designated ``C'', intensities are taken directly from their paper. Direct comparisons between intensities from different measurement methods should not be made.

Columns 3 and 4 list the FWHM (in {\cm}) and S/N for FTS lines. Columns 5 through 7 give the observed wavelength in air and vacuum, and the associated wavelength uncertainty. Air wavelengths are listed for transitions between 200~nm and 2~$\mu$m and are derived from vacuum wavelengths using the five-parameter dispersion formula of \citet{Peck:72}. Observed wavenumbers and their uncertainties are presented in columns 8 and 9. Columns 10 and 11 contain Ritz wavenumbers and uncertainties, calculated from the optimised energy levels determined in this work. Values in columns 5, 6, 8, and 10 have been rounded according to the “rule of 25,” ensuring that the uncertainty in the final digit does not exceed 25.

The upper and lower energy level designations for each transition are given in columns 12 and 13, with their corresponding optimised energy values listed in columns 14 and 15. The final column provides notes on individual lines, including indications of blending, poor line quality, or lines assigned reduced weights during the energy level optimisation.

\begin{splitdeluxetable*}{llllllllllBllllllllll}
\tabletypesize{\scriptsize}
\tablecaption{Classified Lines of \mn \label{tab:linelist}}
\tablehead{\colhead{Source} & \colhead{Int.} & \colhead{FWHM} & \colhead{S/N} & \colhead{$\lambda_{air}$} & \colhead{$\lambda_{vac}$} & \colhead{Unc.} & \colhead{$\sigma_{obs.}$} & \colhead{Unc.} & \colhead{$\sigma_{Ritz}$} & \colhead{Unc.} & 
        \multicolumn{3}{c}{Lower Level} & \multicolumn{3}{c}{Upper Level} & \colhead{$E_L$} & \colhead{$E_U$} & \colhead{Notes} \\
        \cline{12-14} 
        \cline{15-17} 
        \colhead{} & \colhead{} & \colhead{($10^{-3}${\cm})} & \colhead{} & \colhead{(nm)} & \colhead{(nm)} & \colhead{(nm)} & \colhead{(\cm)} & \colhead{(\cm)} & \colhead{(\cm)} & \colhead{(\cm)} & \colhead{Config.} & \colhead{Term} & \colhead{J} & \colhead{Config.} & \colhead{Term} & \colhead{J} & \colhead{(\cm)} & \colhead{(\cm)} & \colhead{}}
\startdata
F & 2.41 & 39 & 6  & - & 51112.337 & 0.015 & 1956.478  & 0.007  & 1956.475  & 0.0005 & $3d^6(^5D)4p$        & $x^6P$ & $\nicefrac{3}{2}$ & $3d^54s(^7S)4d$      & $x^6P$ & $\nicefrac{5}{2}$ & 45259.1735 & 47215.6484 \\
F & 2.62 & 45 & 8  & - & 51046.145 & 0.019 & 1959.000  & 0.005  & 1959.012  & 0.0007 & $3d^6(^5D)4p$        & $x^6P$ & $\nicefrac{3}{2}$ & $3d^54s(^7S)4d$      & $x^6P$ & $\nicefrac{3}{2}$ & 45259.1735 & 47218.1853 \\
F & 2.25 & 14 & 10 & - & 51006.649 & 0.018 & 1960.5277 & 0.0024 & 1960.529  & 0.0007 & $3d^6(^5D)4p$        & $x^6P$ & $\nicefrac{3}{2}$ & $3d^54s(^7S)4d$      & $x^6P$ & $\nicefrac{1}{2}$ & 45259.1735 & 47219.7023 \\
F & 3.09 & 56 & 20 & - & 48638.746 & 0.011 & 2055.9749 & 0.0024 & 2055.974  & 0.0005 & $3d^6(^5D)4p$        & $x^6P$ & $\nicefrac{5}{2}$ & $3d^54s(^7S)4d$      & $x^6P$ & $\nicefrac{7}{2}$ & 45156.1085 & 47212.0826 \\
F & 2.92 & 50 & 15 & - & 48554.536 & 0.012 & 2059.531  & 0.003  & 2059.540  & 0.0005 & $3d^6(^5D)4p$        & $x^6P$ & $\nicefrac{5}{2}$ & $3d^54s(^7S)4d$      & $x^6P$ & $\nicefrac{5}{2}$ & 45156.1085 & 47215.6484 \\
F & 2.89 & 52 & 14 & - & 45082.664 & 0.010 & 2218.145  & 0.003  & 2218.148  & 0.0005 & $3d^6(^5D)4p$        & $x^6P$ & $\nicefrac{7}{2}$ & $3d^54s(^7S)4d$      & $x^6P$ & $\nicefrac{7}{2}$ & 44993.935  & 47212.0826 \\
F & 3.03 & 43 & 23 & - & 42880.46  & 0.03  & 2332.0644 & 0.0019 & 2332.064  & 0.0019 & $3d^54s(^7S)6s$      & $f^8S$ & $\nicefrac{7}{2}$ & $3d^54s(^7S)6p$      & $f^8S$ & $\nicefrac{5}{2}$ & 50157.7535 & 52489.8179 \\
F & 3.18 & 73 & 19 & - & 42744.85  & 0.05  & 2339.464  & 0.003  & 2339.463  & 0.003  & $3d^54s(^7S)6s$      & $f^8S$ & $\nicefrac{7}{2}$ & $3d^54s(^7S)6p$      & $f^8S$ & $\nicefrac{7}{2}$ & 50157.7535 & 52497.217  \\
F & 3.10 & 52 & 37 & - & 42570.38  & 0.03  & 2349.055  & 0.003  & 2349.0511 & 0.0017 & $3d^54s(^7S)6s$      & $f^8S$ & $\nicefrac{7}{2}$ & $3d^54s(^7S)6p$      & $f^8S$ & $\nicefrac{9}{2}$ & 50157.7535 & 52506.8046 \\
F & 2.89 & 37 & 29 & - & 42432.76  & 0.03  & 2356.677  & 0.005  & 2356.6696 & 0.0014 & $3d^54s(^7S)6s$      & $g^6S$ & $\nicefrac{5}{2}$ & $3d^54s(^7S)6p$      & $g^6S$ & $\nicefrac{7}{2}$ & 50904.607  & 53261.2766 \\
F & 2.96 & 22 & 11 & - & 41895.916 & 0.024 & 2386.869  & 0.003  & 2386.8675 & 0.0014 & $3d^54s(^7S)6s$      & $g^6S$ & $\nicefrac{5}{2}$ & $3d^54s(^7S)6p$      & $g^6S$ & $\nicefrac{5}{2}$ & 50904.607  & 53291.4745 \\
F & 2.85 & 59 & 11 & - & 41551.11  & 0.04  & 2406.682  & 0.004  & 2406.6745 & 0.0024 & $3d^54s(^7S)6s$      & $g^6S$ & $\nicefrac{5}{2}$ & $3d^54s(^7S)6p$      & $g^6S$ & $\nicefrac{3}{2}$ & 50904.607  & 53311.282  \\
F & 1.82 & 19 & 3  & - & 37961.394 & 0.016 & 2634.254  & 0.009  & 2634.2552 & 0.0011 & $3d^5(^4P)4s4p(^3P)$ & $v^6P$ & $\nicefrac{3}{2}$ & $3d^54s(^7S)5d$      & $v^6P$ & $\nicefrac{5}{2}$ & 50099.2245 & 52733.4797 \\
F & 2.58 & 63 & 5  & - & 37936.913 & 0.016 & 2635.951  & 0.006  & 2635.9551 & 0.0011 & $3d^5(^4P)4s4p(^3P)$ & $v^6P$ & $\nicefrac{3}{2}$ & $3d^54s(^7S)5d$      & $v^6P$ & $\nicefrac{3}{2}$ & 50099.2245 & 52735.1796 \\
F & 2.72 & 57 & 8  & - & 37923.49  & 0.05  & 2636.884  & 0.008  & 2636.888  & 0.003  & $3d^5(^4P)4s4p(^3P)$ & $v^6P$ & $\nicefrac{3}{2}$ & $3d^54s(^7S)5d$      & $v^6P$ & $\nicefrac{1}{2}$ & 50099.2245 & 52736.112  \\
F & 2.51 & 50 & 6  & - & 37367.797 & 0.008 & 2676.111  & 0.005  & 2676.1011 & 0.0006 & $3d^54s(^7S)4d$      & $e^6D$ & $\nicefrac{7}{2}$ & $3d^5(^4P)4s4p(^3P)$ & $e^6D$ & $\nicefrac{7}{2}$ & 47212.0826 & 49888.1836 \\
F & 2.63 & 46 & 8  & - & 37302.184 & 0.008 & 2680.8065 & 0.0014 & 2680.8082 & 0.0006 & $3d^54s(^7S)4d$      & $e^6D$ & $\nicefrac{9}{2}$ & $3d^5(^4P)4s4p(^3P)$ & $e^6D$ & $\nicefrac{7}{2}$ & 47207.3755 & 49888.1836 \\
F & 3.00 & 23 & 67 & - & 36789.558 & 0.014 & 2718.1636 & 0.0020 & 2718.1626 & 0.0010 & $3d^5(^4P)4s4p(^3P)$ & $v^6P$ & $\nicefrac{5}{2}$ & $3d^54s(^7S)5d$      & $v^6P$ & $\nicefrac{7}{2}$ & 50012.6147 & 52730.7773 \\
F & 2.99 & 42 & 22 & - & 36753.018 & 0.016 & 2720.872  & 0.005  & 2720.8650 & 0.0012 & $3d^5(^4P)4s4p(^3P)$ & $v^6P$ & $\nicefrac{5}{2}$ & $3d^54s(^7S)5d$      & $v^6P$ & $\nicefrac{5}{2}$ & 50012.6147 & 52733.4797 \\
F & 2.84 & 59 & 11 & - & 36730.071 & 0.017 & 2722.562  & 0.016  & 2722.5649 & 0.0013 & $3d^5(^4P)4s4p(^3P)$ & $v^6P$ & $\nicefrac{5}{2}$ & $3d^54s(^7S)5d$      & $v^6P$ & $\nicefrac{3}{2}$ & 50012.6147 & 52735.1796\\ 
\enddata

\tablecomments{\textbf{Note.} The columns are: (1) the source of the line: F - FT spectra, G - grating spectra, C - from \citet{Catalan1964}. (2) relative line intensity. Ft intensity is the log of the relative line intensity. Intensities from the FT and grating spectra are on different scales. Photographic plate line intensities are visual estimates of plate darkening, with qualifiers: w—wide, b—broad, s—shaded to short wavelengths, l—shaded to long wavelengths, p—perturbed (e.g., on a wing of a stronger line), h—hazy, q—questionable position, c—complex (central position is blend of 3 or more lines). (3) full width at half maximum (for FT lines only). (4) signal-to-noise ratio (for FT lines only). (5) Ritz air wavelength - given for lines between 200 nm and 2 $\mu$m. (6) Ritz vacuum wavelength. (7) Ritz wavelength uncertainty. (8) observed wavenumber. (9) observed wavenumber uncertainty. (10) Ritz wavenumber, derived from the optimised upper and lower energy level values of the transition. (11) Ritz wavenumber uncertainty. (12-14) configuration, term and J of the lower energy level. (15-17) configuration, term and J of the upper energy level. (18) lower energy level value. (19) upper energy level value. (20) notes associated with the line: multiply classified with {\mn}, {\mnii} or \ion{Mn}{3} (“I”, “II”, “III”), or another element (e.g. Ar, Ne, Cu). Lines with `\#' were assigned a lower weight in the energy level optimisation. H indicates lines that were fitted with HFS parameters to determine  observed wavenumbers.  (This table is available in its entirety in machine-readable form)}
\end{splitdeluxetable*}

\section{Atomic structure of {\mn}} 
\label{sect:mn_structure}

{\mn} belongs to the iron-group ($3d$) elements and has a ground level configuration of $3d^5 4s^2\,a^6S_{\nicefrac{5}{2}}$. In these elements, the $nd$ and $(n+1)s$ electron configurations have similar binding energies, resulting in significant overlap between the three lowest configurations of {\mn}: $3d^7$, $3d^6 4s$, and $3d^5 4s^2$. This overlap leads to observable two-electron transitions due to configuration interaction among these low-lying states. Within each configuration, strong mixing between LS components allows transitions between normally forbidden intercombination lines, enabling the entire term system of {\mn} to be connected to its ground state.

\begin{figure*}
    \centering
    \includegraphics[width=0.9\linewidth]{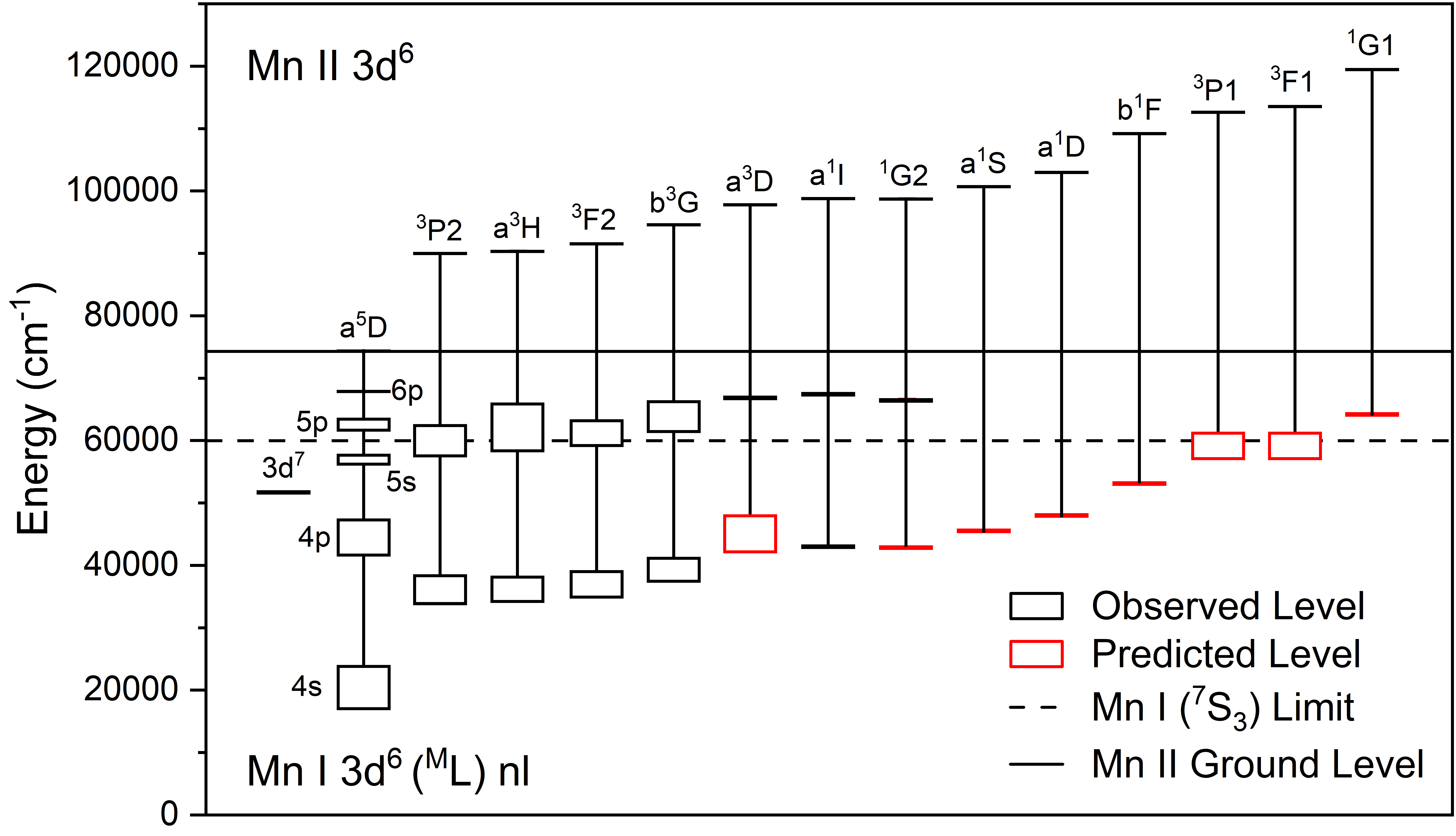}
    \caption{The singly excited system of energy levels in {\mn}.}
    \label{fig:singlyexcited}
\end{figure*}

The energy level structure of {\mn} is divided into two major configuration systems. The singly excited (or ``normal") system consists of one valence electron in the $3d^6(^ML)nl$ subconfigurations, built on the parent terms $3d^6(^ML)$ of {\mnii}. A schematic diagram of this system is shown in Figure~\ref{fig:singlyexcited}, where subconfigurations are grouped by their parent terms from the {\mnii} ground state. Each box in Figure~\ref{fig:singlyexcited} represents the energy range of levels within a given subconfiguration. The $^3F$ and $^3P$ parent terms appear twice in the diagram, differentiated by index numbers. The parent terms $^1H$, $^1F$, and $^3S$ are not included as they are neither known in {\mnii} nor associated with any observed classified transitions in {\mn}.

\begin{figure*}
    \centering
    \includegraphics[width=0.9\linewidth]{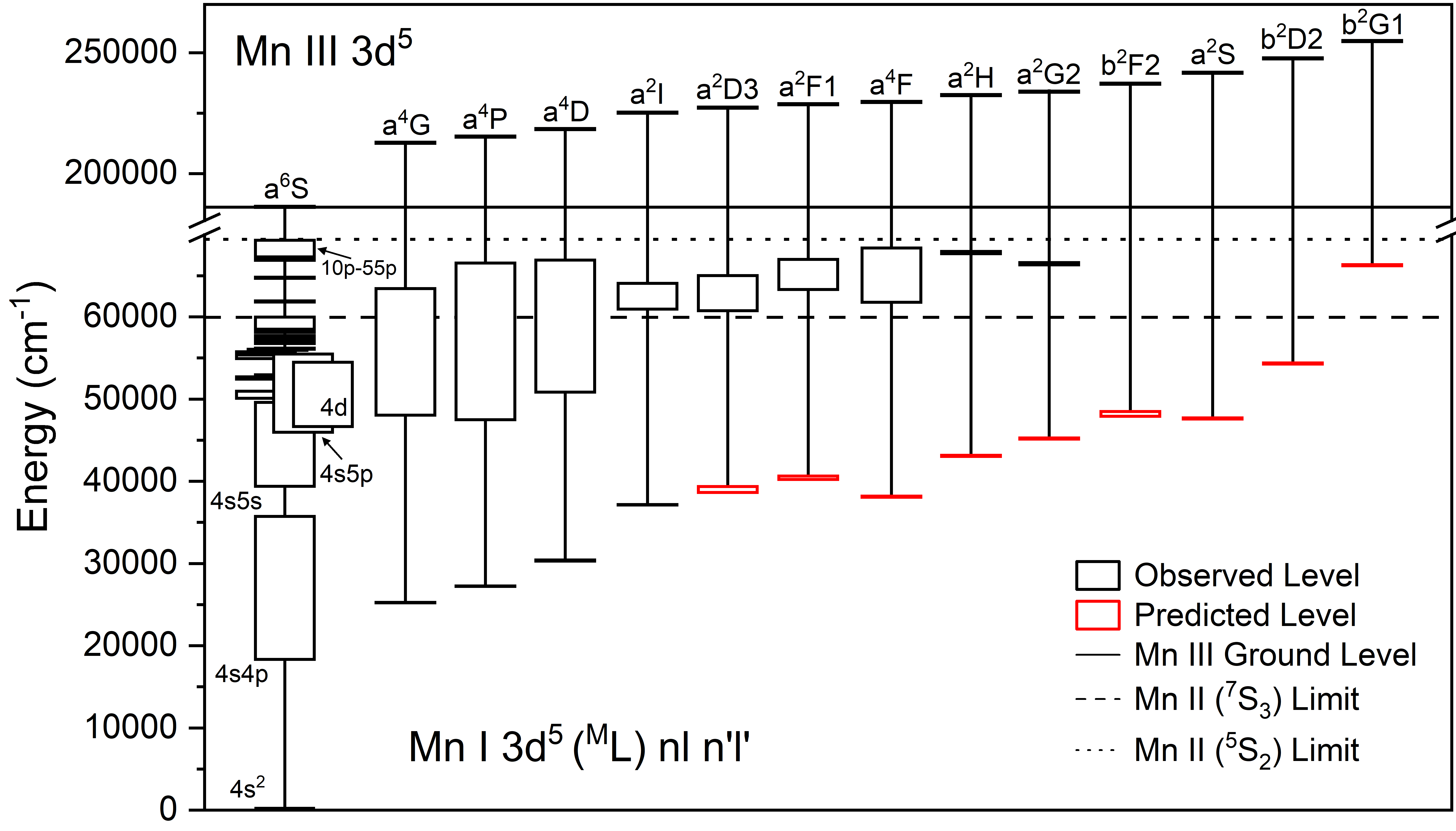}
    \caption{The doubly excited system of energy levels in {\mn}.}
    \label{fig:doublyexcited}
\end{figure*}

\begin{figure}
    \centering
    \includegraphics[width=0.95\linewidth]{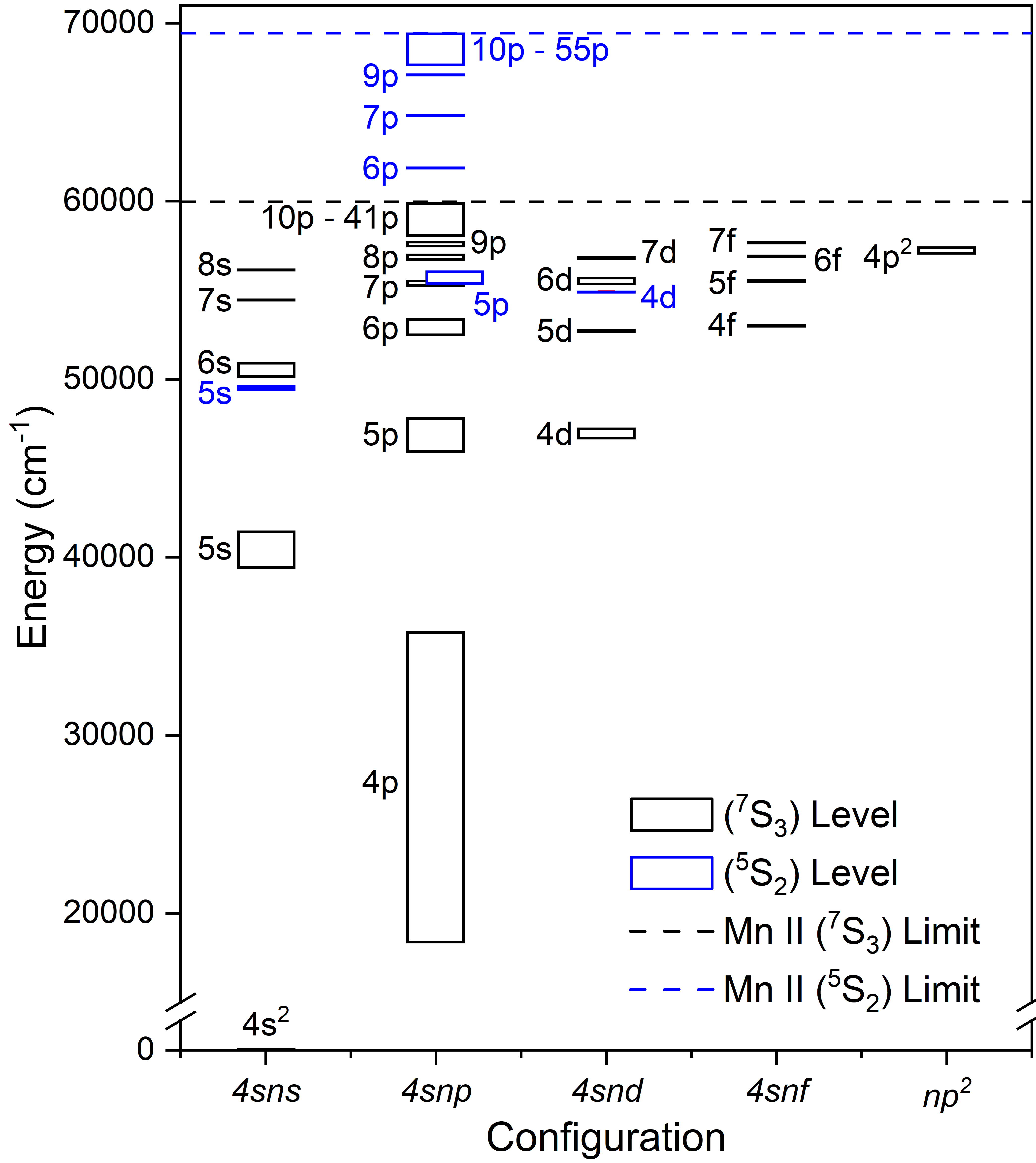}
    \caption{Expanded view of the $3d^5(^6S)nl\,n'l'$ subconfiguration.}
    \label{fig:6S}
\end{figure}

The doubly excited system of {\mn} contains two valence electrons in the $3d^5(^ML)nl\,n'l'$ subconfigurations and is built on the grandparent terms $3d^5(^ML)$ of \ion{Mn}{3}. A schematic diagram of this system is shown in Figure \ref{fig:doublyexcited}. As in the singly excited system, most levels below the $^7S_3$ ionisation limit are known, and a large fraction belong to the $3d^5(^6S)$ grandparent ground state system. In S\&C, this $^6S$ grandparent term is described using three different term labels: $^7S$, $^5S$, and $^6S$. The $3d^5(^6S)4s4p$ configuration is assigned to the $^6S$ grandparent term, while configurations above $3d^54s4p$ are described as $3d^54s(^7S)n'l'$ or $3d^54s(^5S)n'l'$. An expanded schematic diagram of the $(^6S)$ parent term is shown in Figure \ref{fig:6S}. This $(^7S)$ or $(^5S)$ description originates from the level identifications of \citet{Catalan1964}. For clarity and consistency we have retained this nomenclature, and our optimised energy levels are classified using these original term designations. 

In the neutral spectrum of manganese, the doubly excited system dominates, with 314 observed levels assigned to doubly excited configurations. In total, excluding the new levels found in this work, there are 550 previously known levels of {\mn} \citep{Sugar1985a}. The doubly excited system constitutes approximately 65\% of the known levels with the singly excited system accounting for approximately 33\% . The remaining $\sim$2\% of known levels are not classified and have no assigned configuration in previously published work. The majority of these levels are above the $^7S_3$ limit and have considerable mixing with other levels. 

It should be noted that 97 of the previously observed levels in the doubly excited system form part of the Rydberg series for the $^7S_3$ and $^5S_2$ limits. These high order $np$ configurations do not have transitions in a majority of astrophysical spectra and are mainly used in the calculation of the ionisation limit of the ground configurations. These levels were not populated in our laboratory spectra and are not included in this analysis.

\section{Energy Level Optimisation}
\label{sect:level_optimisation}

Observed wavenumbers and their associated uncertainties were used in a least-squares fitting procedure with the program LOPT \citep{LOPT} to confirm and optimise previously known energy levels of {\mn}. Each {\mn} transition was weighted by the inverse square of its total wavenumber uncertainty. Transitions with multiple identifications were initially assigned low weightings (i.e. high uncertainties), ensuring that line parameters were retained in the dataset but did not influence the optimisation of energy level values. These weightings were revised as identifications were resolved.

The optimisation process was iterative. It commenced with a small subset of low-lying odd energy levels that exhibited strong transitions to the ground state. Once optimised, these levels were then used to refine the identification of additional {\mn} transitions in the linelist. The process was then repeated for further subsets of energy levels, progressively expanding the number of identified transitions. At each stage, any observed wavenumber that deviated significantly from its Ritz wavenumber (calculated from fitted energy levels) was investigated for possible misidentification, blending, or HFS. In such cases, the line identification was removed, the weighting reduced, or the spectral line remeasured, respectively. This iterative procedure was continued until all previously known energy levels that could be confirmed were optimised. A detailed search was then performed for new energy levels associated with previously unclassified spectral lines. Details of discovery of new levels are given in Section \ref{sect:new_levels}.

All energy level values, both revised and newly identified, are presented in Table~\ref{tab:levels}. Columns one through three list the configuration, term, and $J$-value of each energy level. Columns four and five give the optimised energy and its uncertainty, relative to the ground state.  Flags indicating special cases (e.g. uncertain identifications, blended levels) are given in column six, with definitions provided in the table footnote. The final column lists the total number of transitions used in the optimisation of each level.

The uncertainty of the energy levels is the uncertainty of the level structure least squares fit from LOPT with the linear addition of the uncertainty of the original \ion{Ar}{2} standard lines (1 part in $10^8$). This is because the uncertainty of the standard lines is common to all observed lines and thus it becomes weighted out of the least squares fit and the resultant level uncertainties during level optimisation. We have adopted a global calibration uncertainty, based on the maximum calibration uncertainty in Table \ref{tab:keff1}, of 3 parts in $10^8$ (from spectrum mmnh). We did not adopt the 5 parts in $10^8$ global calibration uncertainty used in \citet{Liggins2021} which was based on the calibration uncertainty of spectrum ``2012 Sep 13'', as the hollow cathode discharge source for this spectrum was optimised for {\mnii} and only produced a small number of weak {\mn} lines that were used in our analysis. Our global calibration uncertainty was set as the minimum
uncertainty for all optimised energy level values and Ritz wavenumbers.

A discussion of the energy levels listed in S\&C which have been confirmed and optimised, with substantial reductions in uncertainties, following our analysis is given in the following sections. The designations of 52 energy levels have also been redesignated following our analysis and are detailed in Table \ref{tab:new_level_desigs}.

\begin{table*}[h]
    \centering
    \caption{Energy Levels of \mn}
    \begin{tabular}{llllllll}
        \hline\hline
        \multicolumn{4}{c}{Energy Level} & $E$ & Unc. & Flag & No. of \\ \cline{1-4}
        Configuration & Term & J & Parity& ({\cm}) & ({\cm}) &  &  Lines\\ \hline
$3d^54s^2$ & $a^6S$ & $\nicefrac{5}{2}$ & even & 0.0 & 0.0 & & 77 \\
$3d^6(^5D)4s$ & $a^6D$ & $\nicefrac{9}{2}$ & even & 17052.3043 & 0.0005 & & 46 \\
$3d^6(^5D)4s$ & $a^6D$ & $\nicefrac{7}{2}$ & even & 17282.0213 & 0.0005 & & 77 \\
$3d^6(^5D)4s$ & $a^6D$ & $\nicefrac{5}{2}$ & even & 17451.5369 & 0.0005 & & 80 \\
$3d^6(^5D)4s$ & $a^6D$ & $\nicefrac{3}{2}$ & even & 17568.4819 & 0.0005 & & 62 \\
$3d^6(^5D)4s$ & $a^6D$ & $\nicefrac{1}{2}$ & even & 17637.1570 & 0.0005 & & 45 \\
$3d^5(^6S)4s4p(^3P)$ & $z^8P$ & $\nicefrac{5}{2}$ & odd & 18402.4843 & 0.0006 & & 23 \\
$3d^5(^6S)4s4p(^3P)$ & $z^8P$ & $\nicefrac{7}{2}$ & odd & 18531.6651 & 0.0006 & & 26 \\
$3d^5(^6S)4s4p(^3P)$ & $z^8P$ & $\nicefrac{9}{2}$ & odd & 18705.3600 & 0.0007 & & 23 \\
$3d^6(^5D)4s$ & $a^4D$ & $\nicefrac{7}{2}$ & even & 23296.6821 & 0.0007 & & 67 \\
$\cdots$ \\
$3d^5(^2G)4s4p(^3P)$ & $u^2G$ & $\nicefrac{9}{2}$ & odd & 68285.741 & 0.015 & & 5 \\
$3d^5(^2G)4s4p(^3P)$ & $u^2G$ & $\nicefrac{7}{2}$ & odd & 68338.978 & 0.014 & & 3 \\
$3d^5(^4G)4s5s$ & $^4G$ & $\nicefrac{11}{2}$ & even & 68693.082 & 0.003 & & 4 \\
$3d^5(^4G)4s5s$ & $^4G$ & $\nicefrac{9}{2}$ & even & 68716.3459 & 0.0023 & & 5 \\
$3d^6(^1I)4p$ & $z^2K$ & $\nicefrac{15}{2}$ & odd & 68797.703 & 0.013 & & 1 \\
$3d^6(^1I)4p$ & $z^2K$ & $\nicefrac{13}{2}$ & odd & 68842.387 & 0.025 & & 3 \\
$3d^5(^2I)4s4p$ & $x^2I$ & $\nicefrac{13}{2}$ & odd & 69560.811 & 0.01 & & 2 \\
$3d^5(^2I)4s4p$ & $x^2I$ & $\nicefrac{11}{2}$ & odd & 69629.671 & 0.018 & & 3 \\
$3d^6(^1I)4p$ & $v^2H$ & $\nicefrac{9}{2}$ & odd & 69663.163 & 0.011 & & 7 \\
$3d^6(^1I)4p$ & $v^2H$ & $\nicefrac{11}{2}$ & odd & 69722.908 & 0.006 & & 7 \\
\hline \\
    \end{tabular}
    \label{tab:levels}
    \\
    \textbf{Note.} The columns are: (1-3) The configuration, term and J-value of the energy level. (4) The energy level value. (5) Energy level value uncertainty. (6) Flag: N = new level, A = anomalous hyperfine structure. (7) Number of lines contributing to the energy level optimisation.  (This table is available in its entirety in machine-readable form)
\end{table*}

\subsection{The Even Levels of the Singly Excited System}
\label{sect:singly_excited_even}

\subsubsection{The $3d^7$ and $3d^6(^ML)4s$ configurations}
As in S\&C, only the three levels of the $3d^7\;\,^4P$ term have been found for this configuration. There were no lines present in our spectra for transitions to the remaining terms of the $3d^7$ configuration. 

All of the levels of the $3d^6(^ML)4s$ subconfiguration for the $(^5D), (^3P_2), (^3H), (^3F_2), (^3G)$ and $(^1I)$ parent terms in S\&C have been identified and optimised following our analysis, and these subconfigurations are now complete. No transitions to the levels of the remaining parent term $4s$ subconfigurations were present in our spectra, and these levels remain unknown.

\subsubsection{The $3d^6(^5D)5s$ and $3d^6(^5D)4d$ subconfigurations}
No energy levels corresponding to the configurations $3d^6 5s$ or $3d^6 4d$ arising from parent terms higher than $3d^6(^5D)$ are reported in Sugar and Corliss (S\&C). Consistent with this, all levels identified in our analysis for these configurations also originate exclusively from the $3d^6(^5D)$ parent term.
For the $(^5D)5s$ levels, we have assigned the levels of the $e^4D$ term, which were given without configurations in S\&C as $(^5D)5s\,\; e^4D$. The $(^5D)5s\,\; e^4D_{\nicefrac{3}{2}}$, in S\&C at 56601.63 {\cm}, was discarded and replaced by a new level value at 56617.4657 {\cm}. The $(^5D)5s\,\; e^4D$ term remains incomplete with $e^4D_{\nicefrac{1}{2}}$ still unknown. The same is true for the $(^5D)5s\,\; i^6D$ term with with $i^6D_{\nicefrac{1}{2}}$ missing. We have redesignated the levels labelled as $(^5D)5s\,\; f^4D$ in S\&C as $3d^6(^5D)5s\;\,f^4D$ levels. 

For the $(^5D)4d$ subconfigurations, the following levels in S\&C could not be confirmed with the lines present in our FT or grating spectra: $e^4F_{\nicefrac{9}{2},\,\nicefrac{7}{2}}$, $e^6F_{\nicefrac{3}{2}}$ and $e^6G_{\nicefrac{5}{2},\,\nicefrac{3}{2}}$. The remaining S\&C levels of the $(^5D)4d$ subconfiguration have all been confirmed and optimised following our analysis.

\subsection{The Odd Levels of the Singly Excited System}
\label{sect:singly_excited_odd}

\subsubsection{The $3d^6(^ML)4p$ subconfigurations}
In total, 90\% of the levels of the $3d^6(^ML)4p$ subconfigurations listed in S\&C have been confirmed and optimised following our analysis. The $3d^6(^5D)4p$ subconfiguration is complete, as is the $3d^6(^3H)4p$ subconfiguration with levels now assigned to the designations $3d^6(^3H)4p\,\;v^4G_{\nicefrac{5}{2}}$ and $v^4G_{\nicefrac{7}{2}}$ at 65873.2859 {\cm} and 65876.2678 {\cm} respectively. The S\&C level $3d^6(a^3F)4p\,\;^2F_{\nicefrac{5}{2}}$ at 61469.70 {\cm} has been discarded and replaced with a new level found at 61471.2400 {\cm}. 
10 levels were not optimised in our analysis due to the lack of new FT or grating lines in our spectra. These mostly belong to the $3d^6(^3G)4p$ subconfiguration, with 2 levels from the $3d^6(^3D)4p$ subconfiguration and 1 from the $3d^6(^3H)4p$ subconfiguration. These levels are not confirmed in our work and are therefore omitted from Table \ref{tab:levels}.

\subsubsection{The $3d^6(^5D)6p$ subconfiguration}
The only $3d^66p$ level listed in S\&C is $3d^6(^5D)6p$ at 67890.00 {\cm}, but it was given neither a term nor J value in their compilation. No lines to this level were present in our spectra and therefore it was not confirmed in our analysis.

\subsection{The Even Levels of the Doubly Excited System}

\subsubsection{Levels of the $3d^54s^2$ configuration}
The two electrons in the $3d^54s^2$ configuration mean that the $4s$ subshell is closed and so the term structure of the configuration is based on the grandparent terms of $3d^5$. we have found $3d^54s^2$ levels for the first five parent terms: $3d^5(^6S)4s^2,(a^4G),(a^4P), (a^4D)$ and $(a^2I)$ and these are all complete and revised in our work. No spectral lines from $3d^54s^2$ levels of the remaining parent terms were found. 

\subsubsection{Levels of the $3d^54s(^7S)ns$ and $3d^54s(^5S)5s$ subconfigurations}
As discussed in Section \ref{sect:mn_structure}, the $3d^5(^6S)$ grandparent term in {\mn} is described using three separate term labels, with levels beyond $4s4p$ being assigned as $3d^54s(^7S)n'l'$ or $3d^54s(^5S)n'l'$. All of the $ns$ levels for these subconfigurations listed in S\&C have been confirmed and optimised in our analysis. The $3d^54s(^5S)5s$ subconfiguration is complete, but no further levels of the $3d^54s(^5S)ns$ subconfigurations have been found. The $3d^54s(^5S)5s$, $6s$ and $7s$ subconfigurations are also complete, with our new level for $3d^54s(^7S)7s\;\,^8S_{\nicefrac{7}{2}}$ at 54181.0432 {\cm}. For the $3d^54s(^5S)8s$ subconfiguration, we could only confirm the single level $g^8S_{\nicefrac{7}{2}}$, with the $^6S_{\nicefrac{5}{2}}$ level remaining unknown. 

\subsubsection{Levels of the $3d^54s(^7S)nd$ and $3d^54s(^5S)5d$ subconfigurations}
The $3d^54s(^5S)5d\,\;g^6D$ term is complete with all 5 levels newly optimised. The remaining 4 levels of the $^4D$ term were not found in S\&C or in our analysis. Both the $3d^54s(^7S)4d$ and $3d^54s(^7S)5d$ subconfigurations are fully complete, with ten levels found and optimised for each. The $3d^54s(^7S)6d$ subconfiguration is nearly complete, with only the two lowest-J levels of the $g^8D$ term unknown.

The three lowest-J levels of the $3d^54s(^7S)5d\;\,f^8D$ term ($f^8D_{\nicefrac{3}{2}}$, $f^8D_{\nicefrac{5}{2}}$ and $f^8D_{\nicefrac{7}{2}}$) were unresolved in the previous work of S\&C with 52702.48 {\cm} given as the energy value for all three levels. This collective value had relied on low resolution grating spectra recorded in the UV and had been determined from a blended transition to $3d^5(^6S)4s4p(^3P)\,\; z^8P_{\nicefrac{5}{2}}$ identified at 34300.02 {\cm}. This transition is labelled as``wide and hazy" in \citet{Catalan1964} which was attributed to unresolved hyperfine structure. With the high resolution of FTS and the new IR FT spectra, strong, resolved transitions to these levels mean these have now been fully resolved. 
Positive identifications were also made with strong IR lines to $(^7S)5p$ levels, further securing the $f^8D$ levels.

The five $3d^54s(^7S)7d\,\; h^8D_J$ levels were not resolved in S\&C with 56801.4 {\cm} being given as a collective energy. \citet{Catalan1964} had identified 4 transitions in the visible to secure these levels, but these lines were weak (intensity in Catal\'an of 1 and 5) and were labelled as hazy, probably indicating significant HFS which made them unsuitable for the determination of individual energy levels. These lines are resolved in our FTS spectra and enabled the establishment of four separate $h^8D_J$ energy levels, resolved into $J = 11/2$, $9/2$, $7/2$ and $5/2$. Lines to the $J = 3/2$ level were not observed in our new spectra and therefore an energy for this level was not determined.

\subsection{The Odd Levels of the Doubly Excited System}

\subsubsection{Levels of the $3d^5(^2L)4s4p$ and $3d^5(^4L)4s4p$ subconfigurations}

In total, 38 out of 43 $4s4p$ levels belonging to doublet grandparent $3d^5(^2L)$ subconfigurations and 76 out of 86 levels belonging to quartet grandparent $3d^5(^4L)$ subconfigurations listed in S\&C have been confirmed and optimised. Levels for subconfigurations of the highest doublet grandparent terms, $(b^2F), (a^2S), (b^2D)$ and $(b^2G)$ remain unknown with no transitions to these levels observed. 

The $3d^5(^4D)4s4p\,\;^4F$ term is now complete following the redesignation of the levels 55405.1456 {\cm} and 55368.6389 {\cm} as $^4F_{\nicefrac{5}{2}}$ and $^4F_{\nicefrac{3}{2}}$ respectively.

\subsubsection{Levels of the $3d^5(^6S)4s4p$, $3d^54s(^7S)np$ and $3d^54s(^5S)np$ subconfigurations}
The $3d^5(^6S)4s4p$ subconfiguration is complete with 12 levels confirmed and optimised. 

All of the  $3d^54s(^5S)5p$ levels listed in S\&C have also been confirmed. We were only able to determine one level for higher subconfigurations, $3d^54s(^5S)7p\;\,^6P_{\nicefrac{5}{2}}$, which was listed without a J-value in S\&C, but which we were able to determine to be $J=5/2$. 

The same situation is true for the $3d^54s(^7S)np$ subconfigurations with all levels confirmed for the lower lying subconfigurations of $3d^54s(^7S)5p$ and $3d^54s(^7S)6p$. We were unable to determine values for two $7p$ levels, $3d^54s(^7S)7p\,\;^8P_{\nicefrac{7}{2}}$ and $^8P_{\nicefrac{5}{2}}$, and one $8p$ level, $3d^54s(^7S)8p\,\;^8P_{\nicefrac{7}{2}}$. As discussed in Section \ref{sect:mn_structure}, high order $np$ configurations were not populated in our hollow cathode sources and therefore no spectral lines for transitions to subconfigurations above $3d^54s(^7S)8p$ were observed in our spectra.

\subsubsection{Levels of the $3d^54s(^7S)nf$ subconfigurations}
Five levels each of the $w^6F$ and $z^8F$ terms of the $3d^54s(^7S)4f$ subconfiguration have been found. The level $3d^54s(^7S)4f\;\, w^6F_{\nicefrac{1}{2}}$, also not listed in S\&C, was unable to be confirmed from lines in our spectra. In S\&C, all the levels of the $3d^54s(^7S)4f\;\, z^8F$ term were given the energy 52974.50 {\cm}, but again due to the high resolution of our FT spectra, we have resolved the 5 highest-J levels and have been able to determine distinct level values for each. The two remaining, lowest-J levels of $z^8F_J$ could not be found as their transitions are predicted to be weak.

Following our analysis, the term values of the $3d^54s(^7S)5f$ subconfigurations have been interchanged as the transitions observed in our FT and grating spectra matched far better with these exchanged term values. In addition, two levels designated $3d^54s(^7S)5f\;\,v^6F_{\nicefrac{11}{2}}$ and $v^6F_{\nicefrac{9}{2}}$ in S\&C (now redesignated $y^8F_{\nicefrac{11}{2}}$ and $y^8F_{\nicefrac{9}{2}}$ respectively) at 55499.09 {\cm} have been resolved into two separate levels at 55499.709 {\cm} and 55499.757 {\cm}. 

S\&C list levels in the $3d^54s(^7S)6f$ and $7f$ subconfigurations, but we were unable to confirm these levels as many high-lying levels were not populated in our hollow cathode sources and no spectral lines associated with them were present in our spectra.

\subsubsection{Levels of the $3d^54p^2$ subconfiguration}
Only three levels, $e^8P_{\nicefrac{9}{2}}$, $e^8P_{\nicefrac{7}{2}}$ and $e^8P_{\nicefrac{5}{2}}$, of the $3d^54p^2$ subconfiguration, also listed in S\&C, were found and optimised in our analysis.

\subsection{Anomalous hyperfine structure}
During the course of our analysis we noted that the hyperfine levels of several terms could not be fitted due to the dominance of off-diagonal hyperfine interactions. In normal cases, HFS is described using the diagonal interaction constants A and B, which assume that the total angular momentum quantum number $J$ is a good quantum number and that mixing between fine structure levels is negligible \citep{Ding2020}. However, the $z^8F, f^8D, x^6D$ and $w^6F$ terms exhibit strong hyperfine interactions, whilst the fine structure splitting between individual levels is unusually small. This combination leads to significant mixing between hyperfine levels derived from different fine structure components, rendering the diagonal approximation invalid and introducing large intensity redistributions among the component lines \citep{Andersson2015}. As a result, $J$ loses its physical meaning, and the energy shifts and line intensities cannot be accurately described using standard hyperfine constants. Resolving these transitions would require extensive theoretical calculations accounting for the full off-diagonal interaction matrix for each transition which are beyond the scope of the present work. Levels which are potentially affected by anomalous hyperfine interaction have been highlighted in Table \ref{tab:levels} with ``A'', to indicate the breakdown of the conventional hyperfine description and highlight that care should be taken when using these levels to determine Ritz wavelengths for transitions. 

\subsection{Summary of the Revision of Known Energy Level Values}
\label{sect:prev_levels}

\begin{figure*}
    \centering
    \includegraphics[width=0.8\linewidth]{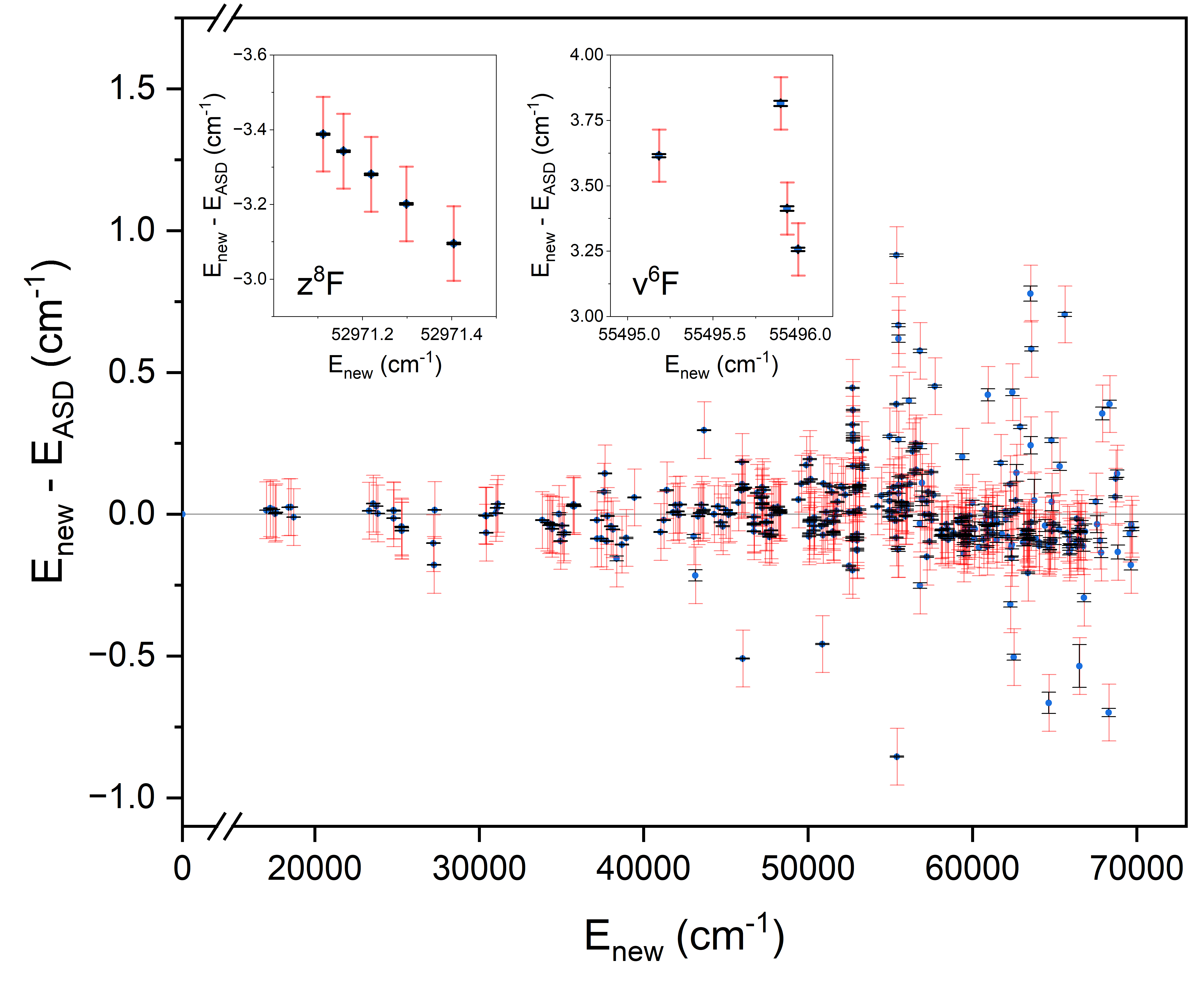}
    \caption{Comparison of our revised {\mn} energy levels and the previously published values of \citet{Sugar1985a}. The uncertainties of S\&C are shown in red and our uncertainties are in black. The levels of the $z^8F$ and $v^6F$ terms have now been resolved and shifted considerably in our new analysis (note the different axes scales). These levels are given in inset plots to ensure that the main plot scale provides sufficient detail of the other levels.}
    \label{fig:E_comp}
\end{figure*}

In total, 384 of the 453 energy levels given in S\&C (excluding Rydberg series levels) have been revised and optimised as a result of this analysis. The energy differences between our new values and those of S\&C are shown in Figure \ref{fig:E_comp}. Error bars show the uncertainties of the energy levels from this analysis (in black) and from S\&C's analysis of previously published levels (in red). The substantial reduction in uncertainty from the present analysis can be seen. Energy levels with the highest precision are typically those belonging to the quartet and sextet terms, due to their strong LS-allowed transitions to the ground state. 
Levels which were optimised exclusively using grating lines may have uncertainties exceeding 0.02 {\cm} due to the much higher wavelength uncertainties compared to FTS measurements. In addition, the level designations of 52 levels have been revised following our analysis. These changes to energy level labels are detailed in Table ~\ref{tab:new_level_desigs}.

\startlongtable
\begin{deluxetable*}{lrr}
\tablewidth{0pt} 
\tablecaption{{\mn} Energy Level Label Redesignations \label{tab:new_level_desigs}}
\tablehead{
        \colhead{Energy} & \colhead{Previous Label} & \colhead{New Label} \\
        \colhead{({\cm})} & \colhead{\citep{Sugar1985a}} & \colhead{(this work)} 
        }
\startdata
55368.6389 & $3d^54s(^5S)5p\;\,w^4P_{\nicefrac{3}{2}}$ & $3d^5(^4D)4s4p\;\,^4F_{\nicefrac{3}{2}}$ \\
55405.1456 & $3d^54s(^5S)5p\;\,w^4P_{\nicefrac{5}{2}}$ & $3d^5(^4D)4s4p\;\,^4F_{\nicefrac{5}{2}}$ \\
55495.185  & $3d^54s(^7S)5f\;\,v^6F_{\nicefrac{5}{2}}$ & $3d^54s(^7S)5f\;\,y^8F_{\nicefrac{5}{2}}$ \\
55495.895  & $3d^54s(^7S)5f\;\,v^6F_{\nicefrac{7}{2}}$ & $3d^54s(^7S)5f\;\,y^8F_{\nicefrac{7}{2}}$ \\
55495.933  & $3d^54s(^7S)5f\;\,v^6F_{\nicefrac{9}{2}}$ & $3d^54s(^7S)5f\;\,y^8F_{\nicefrac{9}{2}}$ \\
55495.997  & $3d^54s(^7S)5f\;\,v^6F_{\nicefrac{11}{2}}$ & $3d^54s(^7S)5f\;\,y^8F_{\nicefrac{11}{2}}$ \\
55499.709  & $3d^54s(^7S)5f\;\,y^8F_{\nicefrac{11}{2}}$ & $3d^54s(^7S)5f\;\,v^6F_{\nicefrac{11}{2}}$ \\
55499.757  & $3d^54s(^7S)5f\;\,y^8F_{\nicefrac{9}{2}}$ & $3d^54s(^7S)5f\;\,v^6F_{\nicefrac{9}{2}}$ \\
55499.764  & $3d^54s(^7S)5f\;\,y^8F_{\nicefrac{7}{2}}$ & $3d^54s(^7S)5f\;\,v^6F_{\nicefrac{7}{2}}$ \\
55499.778  & $3d^54s(^7S)5f\;\,y^8F_{\nicefrac{5}{2}}$ & $3d^54s(^7S)5f\;\,v^6F_{\nicefrac{5}{2}}$ \\
55499.85   & $3d^54s(^7S)5f\;\,y^8F_{\nicefrac{3}{2}}$ & $3d^54s(^7S)5f\;\,v^6F_{\nicefrac{3}{2}}$ \\
56462.2247 & $\;\,e^4D_{\nicefrac{7}{2}}$ & $3d^6(^5D)5s\;\,e^4D_{\nicefrac{7}{2}}$ \\
56562.1994 & $\;\,e^4D_{\nicefrac{5}{2}}$ & $3d^6(^5D)5s\;\,e^4D_{\nicefrac{5}{2}}$ \\
56617.4657 & $\;\,e^4D_{\nicefrac{3}{2}}$ & $3d^6(^5D)5s\;\,e^4D_{\nicefrac{3}{2}}$ \\
57305.6987 & $3d^6(^5D)5s\;\,f^4D_{\nicefrac{7}{2}}$ & $3d^64s(^5S)4p\;\,^4D_{\nicefrac{7}{2}}$ \\
57486.1186 & $3d^6(^5D)5s\;\,f^4D_{\nicefrac{5}{2}}$ & $3d^64s(^5S)4p\;\,^4D_{\nicefrac{5}{2}}$ \\
57621.968  & $3d^6(^5D)5s\;\,f^4D_{\nicefrac{3}{2}}$ & $3d^64s(^5S)4p\;\,^4D_{\nicefrac{3}{2}}$ \\
57706.281  & $3d^6(^5D)5s\;\,f^4D_{\nicefrac{1}{2}}$ & $3d^64s(^5S)4p\;\,^4D_{\nicefrac{1}{2}}$ \\
59989.811  & $3d^6(a^3P)4p\;\,^4D_{\nicefrac{3}{2}}$ & $3d^6(a^3P)4p\;\,w^4D_{\nicefrac{3}{2}}$ \\
60141.928  & $3d^6(a^3P)4p\;\,^4D_{\nicefrac{1}{2}}$ & $3d^6(a^3P)4p\;\,w^4D_{\nicefrac{1}{2}}$ \\
60395.523  & $3d^6(a^3P)4p\;\,^2D_{\nicefrac{3}{2}}$ & $3d^6(a^3P)4p\;\,z^2D_{\nicefrac{3}{2}}$ \\
61469.12   & $3d^6(^3G)4p\;\,^4G_{\nicefrac{11}{2}}$ & $3d^6(^3G)4p\;\,w^4G_{\nicefrac{11}{2}}$ \\
62670.96   & $3d^5(^4F)4s4p(^3P^)\;\,^6D_{\nicefrac{9}{2}}$ & $3d^5(^4F)4s4p(^3P)\;\,w^6D_{\nicefrac{9}{2}}$ \\
63139.604  & $3d^5(^4G)4s4p(^1P^)\;\,^4F_{\nicefrac{5}{2}}$ & $3d^5(a^2F)4s4p\;\,^4G_{\nicefrac{5}{2}}$ \\
63288.7081 & $3d^6(^3G)4p\;\,^2H_{\nicefrac{11}{2}}$ & $3d^6(^3G)4p\;\,z^2H_{\nicefrac{11}{2}}$ \\
63288.714  & $3d^5(^4G)4s4p(^1P^)\;\,_{\nicefrac{7}{2}}$ & $3d^5(a^2F)4s4p\;\,^4G_{\nicefrac{7}{2}}$ \\
63548.4622 & $3d^6(^3G)4p\;\,^2H_{\nicefrac{9}{2}}$ & $3d^6(^3G)4p\;\,z^2H_{\nicefrac{9}{2}}$ \\
64051.7993 & $3d^5(^2I)4s4p(^3P^)\;\,^2I_{\nicefrac{11}{2}}$ & $3d^5(^2I)4s4p\;\,^2I_{\nicefrac{13}{2}}$ \\
64055.271  & $\;\,_{\nicefrac{9}{2}}$ & $3d^5(^2I)4s4p\;\,^2I_{\nicefrac{11}{2}}$ \\
64585.331  & $3d^5(a^2F)4s4p(^3P^)\;\,^4F_{\nicefrac{9}{2}}$ & $3d^5(a^2F)4s4p\;\,x^2G_{\nicefrac{9}{2}}$ \\
64649.082  & $3d^5(a^2F)4s4p(^3P^)\;\,w^2G_{\nicefrac{7}{2}}$ & $3d^5(a^2F)4s4p\;\,x^2G_{\nicefrac{7}{2}}$ \\
64806.961  & $3d^54s(^5S)7p\;\,^6P_{}$ & $3d^54s(^5S)7p\;\,^6P_{\nicefrac{5}{2}}$ \\
65305.299  & $\;\,_{\nicefrac{7}{2}}$ & $3d^5(a^2F)4s4p(^3P)\;\,w^2G_{\nicefrac{7}{2}}$ \\
65617.305  & $3d^6(^3G)4p\;\,_{\nicefrac{7}{2}}$ & $3d^6(^3G)4p\;\,v^2F_{\nicefrac{7}{2}}$ \\
65768.744  & $3d^6(^3G)4p\;\,_{\nicefrac{9}{2}}$ & $3d^6(^3G)4p\;\,^2G_{\nicefrac{9}{2}}$ \\
65873.286  & $3d^6(^3H)4p\;\,_{\nicefrac{5}{2}}$ & $3d^6(^3H)4p\;\,v^4G_{\nicefrac{5}{2}}$ \\
65876.268  & $3d^6(^3H)4p\;\,_{\nicefrac{7}{2}}$ & $3d^6(^3H)4p\;\,v^4G_{\nicefrac{7}{2}}$ \\
65961.777  & $3d^5(^4P)4s4p(^1P^)\;\,_{\nicefrac{3}{2}}$ & $3d^5(^2D)4s4p\;\,w^2D_{\nicefrac{3}{2}}$ \\
66395.077  & $3d^5(^4F)4s4p(^3P^)\;\,_{\nicefrac{5}{2}}$ & $3d^5(^4F)4s4p(^3P)\;\,u^4G_{\nicefrac{5}{2}}$ \\
66454.18   & $3d^5(^4F)4s4p(^3P^)\;\,_{\nicefrac{7}{2}}$ & $3d^5(^4F)4s4p(^3P)\;\,u^4G_{\nicefrac{7}{2}}$ \\
67504.945  & $3d^6(^1I)4p\;\,w^2H_{\nicefrac{11}{2}}$ & $3d^6(^1G)4p\;\,w^2H_{\nicefrac{11}{2}}$ \\
67576.81   & $3d^6(^1I)4p\;\,w^2H_{\nicefrac{9}{2}}$ & $3d^6(^1G)4p\;\,w^2H_{\nicefrac{9}{2}}$ \\
68285.741  & $3d^5(^4F)4s4p(^3P^)\;\,^4F_{\nicefrac{9}{2}}$ & $3d^5(^2G)4s4p(^3P)\;\,u^2G_{\nicefrac{9}{2}}$ \\
68338.978  & $3d^5(^4F)4s4p(^3P^)\;\,^4F_{\nicefrac{7}{2}}$ & $3d^5(^2G)4s4p(^3P)\;\,u^2G_{\nicefrac{7}{2}}$ \\
68693.082  & $3d^54s(^5G)5s\;\,f^4G_{\nicefrac{11}{2}}$ & $3d^5(^4G)4s5s\;\,^4G_{\nicefrac{11}{2}}$ \\
68716.3459 & $3d^54s(^5G)5s\;\,f^4G_{\nicefrac{9}{2}}$ & $3d^5(^4G)4s5s\;\,^4G_{\nicefrac{9}{2}}$ \\
68797.703  & $\;\,z^2K_{\nicefrac{15}{2}}$ & $3d^6(^1I)4p\;\,z^2K_{\nicefrac{15}{2}}$ \\
68842.387  & $\;\,z^2K_{\nicefrac{13}{2}}$ & $3d^6(^1I)4p\;\,z^2K_{\nicefrac{13}{2}}$ \\
69560.811  & $\;\,x^2I_{\nicefrac{13}{2}}$ & $3d^5(^2I)4s4p\;\,x^2I_{\nicefrac{13}{2}}$ \\
69629.671  & $\;\,x^2I_{\nicefrac{11}{2}}$ & $3d^5(^2I)4s4p\;\,x^2I_{\nicefrac{11}{2}}$ \\
69663.163  & $\;\,v^2H_{\nicefrac{9}{2}}$ & $3d^6(^1I)4p\;\,v^2H_{\nicefrac{9}{2}}$ \\
69722.908  & $\;\,v^2H_{\nicefrac{11}{2}}$ & $3d^6(^1I)4p\;\,v^2H_{\nicefrac{11}{2}}$ \\
\enddata
\end{deluxetable*}

\section{Identification of new energy levels}
\label{sect:new_levels}

Following the revision of known energy levels, unclassified lines remaining in our linelist were used to search for new energy levels. To identify potential new energy levels, we combined the wavenumbers of unassigned spectral lines with our optimised energy level dataset. The resulting values were then examined for clusters of transitions falling within a given wavenumber tolerance, typically within $\pm0.05$ {\cm}, that might signify the presence of new levels. Theoretical calculations provided useful guides for approximate level energies and predicted transition strengths to help determine identifications of newly found levels. In particular, the work of the late \citet{Kurucz2016} was invaluable in our search for new levels. In total, 18 new energy levels of {\mn}, reported here for the first time, have been identified and optimised as a result of our analysis. Details of the determination of these new levels are given below.

\subsection{New levels of the $3d^54s(^7S)nl$ subconfigurations} 

\subsubsection{The $3d^54s(^7S)6p\,\; ^8P_{\nicefrac{9}{2}}$ level}
This level was the only level missing from the $4s(^7S)6p\;\, ^8P$ term in S\&C. Our new IR FT measurements, lines to $6s$ and $4d$ levels of the same parent term enabled the optimisation of this level. Three FT lines and one grating line have now been newly classified as transitions from the $3d^54s(^7S)6p\, ^8P_{\nicefrac{9}{2}}$ level at 52506.8046 {\cm}.

\subsubsection{The $3d^54s(^7S)7s\,\; ^8S_{\nicefrac{7}{2}}$ level}
This is the only level within the $3d^54s(^7S)7s\;\,^8S$ term that we were able to determine, as the remaining levels have lines predicted to be very weak and which were not observed in our spectra. The new level at 54181.0432~{\cm} was identified using FT lines in the IR to levels of $4s(^7S)5p$ and in the visible to levels of $(^6S)4s4p(^3P)$. Six new lines have now been classified as transitions from this level.

\subsubsection{The $3d^54s(^7S)5f\,\; y^8F_{\nicefrac{13}{2}}$ level}
A level with this designation had previously been reported in S\&C at 55498.5 {\cm}. However, following our analysis it was determined that the $y^8F$ and $v^6F$ terms should be swapped and that this level energy should be discarded. Our new energy value for $y^8F_{\nicefrac{13}{2}}$ is at 55496.0799 {\cm}, and there are reliable FT transitions in the IR linking to $3d^54s(^7S)4d$ and $5d$ levels, securing this level well in the optimised fit.

\subsubsection{The $3d^54s(^7S)7p\, ^6P_{\nicefrac{3}{2}}$ level}
A new FT transition in the IR, at 6082.3287 {\cm}, enabled the identification of this new $3d^54s(^7S)7p\;\, ^6P_{\nicefrac{3}{2}}$ level, and a grating line in the visible ensured its optimised fit at 55497.7276 {\cm}. This new level completes the $3d^54s(^7S)7p\;\,^6P$ term.

\subsection{The $3d^6(^5D)5s\;\,e^4D_{\nicefrac{9}{2}}$ level}
The previous level in S\&C with the designation $3d^6(^5D)5s\;\,e^4D_{\nicefrac{9}{2}}$ at 56601.63 {\cm} has now been discarded following our analysis. Strong transitions in the visible to levels of the $(^6S)4s4p(^3P)$ subconfiguration enabled the identification of a new level at 56617.4657 {\cm}, and these lines fitted very clearly with theoretical line strengths. In total, five FT and one grating line have been classified for the first time.

\subsection{The $3d^6(^3H)4p\,\;z^4I_{\nicefrac{11}{2}}$ level}
Again, this is a designation that had previously been identified in S\&C, but whose level has now been discarded following our analysis. A new energy level at 58827.4893 {\cm} has determined for this $3d^6(^3H)4p\,\;z^4I_{\nicefrac{11}{2}}$ designation. The strong FT line to $3d^6(^3H)4s\,\;a^4H_{\nicefrac{9}{2}}$ fits very clearly with the multiplet pattern of transitions and relative line intensities of the other $3d^6(^3H)4p\;\,z^4I_J$ levels, providing certainty about the identification of this level.

\subsection{New levels of the quartet parent term $3d^5(^4L)4s4p$ subconfigurations} 
The majority of the new energy levels determined in this work are odd levels of the $4s4p$ subconfigurations with quartet parent terms, $(^4L)$. These are:
\begin{itemize}
    \item $(^4G)4s4p$: $^2H_{\nicefrac{11}{2}}$, $^2H_{\nicefrac{9}{2}}$, $^2F_{\nicefrac{7}{2}}$ and $^2F_{\nicefrac{5}{2}}$
    \item $(^4P)4s4p$: $^2P_{\nicefrac{3}{2}}$, $^2P_{\nicefrac{1}{2}}$ and $^2D_{\nicefrac{3}{2}}$
    \item $(^4D)4s4p$: $^4F_{\nicefrac{9}{2}}$, $^4F_{\nicefrac{7}{2}}$ and $^2F_{\nicefrac{5}{2}}$
    \item $(^4G)4s4p$: $^4G_{\nicefrac{7}{2}}$ and $^2G_{\nicefrac{9}{2}}$
\end{itemize}

It is surprising that these levels were not identified by \citet{Catalan1964} as their transitions are strong and in the visible spectral region where Catal\'an had spectra. It is therefore likely that these levels were weakly populated in the sources used by Catal\'an and the resultant lines in their spectra were very weak. Regardless, these levels were well populated in our sources, and our FT and grating data enabled excellent fits for these new levels to well-secured lower energy levels, resulting in the establishment of 12 new levels and the new classification of 103 spectral lines. 

\subsection{The $(^2I)4s4p\,\; ^4K_{\nicefrac{11}{2}}$ level}
The new $(^2I)4s4p\,\; ^4K_{\nicefrac{11}{2}}$ level is now found and is secured by FT lines in the visible to the singly excited system through transitions to levels of $(^4H)4s$, and to the doubly excited system through transitions to levels of  $3d^54s^2$. Five FT lines have been classified for the first time due to the determination of this level.

\section{Summary}
\label{sect:summary}

New atomic data for {\mn} are urgently needed for the analysis of astrophysical spectra. Previous atomic data for energy levels and wavelength of {\mn} were based on measurements taken over 60 years ago, pre-dating the development of high-resolution FT spectroscopy. These data are now of insufficient accuracy for modern astrophysical spectral analyses.

Following an extensive term analysis of the spectrum of {\mn}, measured using high-resolution FT and grating spectroscopy, 384 previously reported energy levels have been revised with a significant improvement in accuracy - an improvement in uncertainty of at least an order-of-magnitude. 18 completely new {\mn} energy levels have also been found and optimised for the first time. The classified linelist for {\mn} now contains 2186 lines, with 1642 FT and 289 newly-measured grating lines.

This work represents the most accurate {\mn} wavelength and energy level values available to date. These data will enable far more accurate and reliable analyses of {\mn} transition lines in astrophysical spectra.

\begin{acknowledgments}
CPC and JCP thank the STFC of the UK for their support through the grants ST/K001051/1, ST/S000372/1, ST/W000989/1 and UKRI1188. JCP thanks the Royal Society for URF support during the 2001 spectral measurements. In addition, RBW expresses his gratitude to Sveneric Johansson and Hampus Nilsson  for their advice and guidance with term analysis, and gratefully acknowledges the Valery Myerscough Prize for funding his research visit to the Lund Observatory, Sweden. MTB thanks the Ministerio de Ciencia, Innovaci\'on y Universidades of the Spanish Government for her Beatriz Galindo Fellowship and acknowledges funding from MICIU/AEI /10.13039/501100011033 and by FEDER (UE) under project PID2021-127786NA-100. 

JCP thanks the late Bob Kurucz for his extensive theoretical calculations that have aided spectrum analysis research at Imperial over the past 35 years.
\end{acknowledgments}






\bibliography{mn1}{}
\bibliographystyle{aasjournalv7}



\end{document}